# Numerical investigation of airborne transmission in low ceiling rooms under displacement ventilation

Changchang Wang[1], Jiarong Hong[*2]

[1] *Department of Civil and Environmental Engineering, The Hong Kong Polytechnic University, Hung Hom, Kowloon, Hong Kong, China*

[2] *Department of Mechanical Engineering and St. Anthony Falls Laboratory, University of Minnesota, Minneapolis, Minnesota 55455, USA*

## Abstract:

This study employs computational fluid dynamics (CFD) simulations to evaluate the risk of airborne transmission of COVID-19 in low-ceiling rooms, such as elevator cabins, under mechanical displacement ventilation. The simulations take into account the effects of the human body's thermal environment and respiratory jet dynamics on the transmission of pathogens. The results of the study are used to propose a potential mitigation strategy based on ventilation thermal control to reduce the risk of airborne transmission in these types of enclosed indoor spaces. Our findings demonstrate that as the ventilation rate ($Q_v$) increases, the efficiency of removing airborne particles ($\varepsilon_p$) initially increases rapidly, reaches a plateau ($\varepsilon_{p,c}$) at a critical ventilation rate ($Q_c$), and subsequently increases at a slower rate beyond $Q_c$. The $Q_c$ for low-ceiling rooms is lower compared to high-ceiling rooms due to the increased interaction between the thermal plume generated by the occupants or infectors and the ventilation. Further analysis of the flow and temperature fields reveals that $\varepsilon_p$ is closely linked to the thermal stratification fields, as characterized by the thermal interface height, the height of the temperature isosurface, and temperature gradient. The simulations also indicate that the location of infector relative to ventilation inlet/outlet affects $Q_c$ and $\varepsilon_{p,c}$ with higher $Q_c$ and lower $\varepsilon_{p,c}$ observed when infector is in a corner due to potential formation of a local hot spot of high infection risk when infector is near the ventilation inlet. In conclusion, based on the simulations, we propose a ventilation thermal control strategy, by increasing the ventilation temperature, to reduce the risk of airborne transmission in low-ceiling rooms. Our findings indicate that the thermal environment plays a critical role in the transmission of airborne diseases confined spaces.

## I. INTRODUCTION

Airborne transmission is a major transmission pathway that leads to the rapid spread of COVID-19 (SARS-CoV-2 virus) [1-4], particularly in confined crowded indoor spaces. As shown in the literature [5-7], room ventilation plays a crucial role in reducing airborne infection risk, and has been employed as one of the most efficient measures to mitigate the infection risk during the current pandemic [8, 9]. Ventilation has two main modes [5, 6, 10-12], i.e., mixing ventilation (MV) and displacement ventilation (DV). For mixing ventilation, air is circulated throughout the spaces, resulting in a relatively uniform distribution of temperature and particulate matter [10]. For displacement ventilation, air is displaced from the bottom to the top of the room, establishing an internal stratification of temperature and particulate matter [6, 13]. The displacement ventilation mode is usually considered to be better than mixing ventilation in terms of reducing the risk of airborne transmission but may yield lockup phenomena, i.e., the trapping and accumulation of respiratory contaminants between the stratified air layers [14-17]. The effect of such lockup phenomena on airborne transmission is particularly strong for low-ceiling rooms (< 2.45 m) [18], which have been shown to be associated with higher infection risk. Specifically, even before the COVID-19 pandemic, a study of institutional transmission of airborne infections in eight hospitals in Lima, Peru, showed that stuffy, low-ceiling rooms could increase the risk of nosocomial airborne disease transmission compared with large open-windowed, high-ceilinged rooms [19]. In the nosocomial COVID break reported at United Christian Hospital, Hong Kong [20], the low ceiling height was also suggested as one of the contributing factors. In addition to hospitals, a large number of airborne transmissions have been shown to occur in low-ceiling and confined spaces with poor ventilation [21], such as cafeterias [22], aircraft cabins [23], enclosed buses [24], and elevator cabins [25]. Therefore, an in-depth understanding of the lockup phenomena and associated airborne transmission in low ceiling rooms under displacement ventilation is urgently needed.

---
[*] Corresponding author: jhong@umn.edu

Investigation of the effects of thermally stratified lock-up phenomena on airborne transmission by exhaled aerosol particles/droplets under displacement ventilation in indoor environments [26] has been conducted using full-scale experiments and theoretical analysis as well as numerical simulations. To date, full-scale experimental studies are generally conducted using model rooms [13, 27, 28], hospital wards [16] and airborne infection isolation rooms (AIIRs) [29] that are equipped with life-sized breathing thermal manikins inside. For example, Bjørn & Nielsen [13] used photoacoustic spectroscopy to investigate airborne transmission in a model room with a ceiling height of 2.50 m. Their study provided measurements of the vertical contaminant distribution under various heat source intensities and indicated that contaminants exhaled by a breathing manikin could be locked in a thermally stratified layer if the vertical temperature gradient was sufficiently large (above 0.4~0.5 °C/m), resulting in locally high contaminant concentrations. In another full-scale test model room with a high ceiling (2.70 m), Nielsen et al. [27] studied the cross-infection risk between two breathing thermal manikins and found that the exposure increases with decreasing distance between people, with the highest exposure value possibly being up to 12 times in a face-to-face position. In a full-scale hospital ward (ceiling height 2.50 m) with two bed-lying manikins equipped with different ventilation systems, Qian et al. [16] found that compared with mixing and downward ventilation, exhaled gaseous or fine particles could be effectively removed with the assistance of a body plume when facing upward, while exhaled droplet nuclei could accumulate due to thermal stratification lock-up effects when facing sideward. Recently, Jurelionis et al. [28] experimentally investigated aerosol particle dispersion and removal in a full-scale test chamber with high ceilings of 2.80 m and revealed that with increasing ventilation rates, the particle removal efficiency increases in displacement ventilation and is higher than that in mixing ventilation. However, the maximum ventilation rate examined is only 4 ACH (air change rate per hour) in their studies, which is significantly less than the recommended 9 ACH for the free of airborne infection [9]. Berlanga et al. [29] experimentally investigated airborne transmission in a full-scale airborne infection isolation room with high ceilings of 2.80 m, showing that although "lock-up" phenomena could cause a high contaminant concentration zone, the zone influenced by the thermal convective layer around HW shows relatively low exposure. Overall, these experimental studies all suggest a close relationship between airborne transmission in displacement ventilation and thermal stratification lock-up behavior; nevertheless, the underlying physical process involved is not fully understood. In addition, the aforementioned studies are primarily conducted in indoor environments with relatively high ceilings; as a result, the mechanism underlying airborne transmission is still unknown for spaces with low ceilings that are frequently used in daily life (i.e., elevator cabins, cafeterias, *etc.*), where the flow interaction between a thermal plume generated by a heat source and ventilation may be stronger. Several theoretical studies have been performed on the dispersion of airborne respiratory contaminants in indoor environments [30-32]. Zhou et al. [17] developed a theoretical buoyant jet dispersion model in a thermally stratified environment and found a power law relation between the lock-up height and temperature gradient. Although the knowledge of airborne transmission in displacement ventilation is still limited, the lock-up effect and its physics in both nature and mechanical displacement ventilation without considering airborne transmission have been extensively studied. Sandberg & Lindstrom [33] first discussed the stable stratified interface in natural displacement ventilation with a buoyancy source in an enclosure. Linden, Lane-Serff & Smeed [34] theoretically investigated this phenomenon and established the relationship between the interface height ($h$) and effective area of the openings. This work further highlighted the importance of the types and configurations of heat or buoyancy sources on stratification. Subsequently, Cooper & Linden [35] further extended the work of flow and stratification with one single source of buoyancy to that with two heat sources, showing that the height of multiple layers created by multiple sources depends on the height and strength ratio of the two heat sources. For mechanical displacement ventilation driven by mechanical extraction or wind, Hunt & Linden [36-38] established the widely-used formula for quantifying the interface height ($h$) as a function of the total ventilation rate ($Q$) and the buoyancy flux ($B$) generated by heat sources. This formula could provide some fundamental guidance for developing optimal ventilation control to reduce the risk of airborne infection. However, these analytical studies involve substantial simplifications, and the implementation of their findings in a realistic indoor environment requires substantial empirical correction.

Compared with experimental and theoretical studies, Computational Fluid Dynamics (CFD) studies can provide full physical details of flow fields and particle transport (i.e., field distributions of pressure, temperature, velocity, and concentrations of contaminants or particle trajectories). With the advancement in computational power, the CFD method has been extensively utilized to predict and design ventilation flows to prevent airborne transmission in buildings [39-45]. For example, Zhao & Chao [39] used the discrete trajectory model to study airborne transmission in a full-scale room and found that a displacement ventilated room has a larger number of escaped particles and a higher average particle concentration due to its lower deposition rate than the mixing room. However, in this work, no internal source of buoyancy (i.e., occupant) is studied. Gao et al. [42] investigated the dispersion of exhaled droplets ($0.1-10\ \mu m$ in diameter) in an office room with a ceiling height of 2.70 m and demonstrated the importance of contaminant release location (mouth or nose) on the lock-up layer of droplets under displacement ventilation. However, the influence of the ceiling height and ventilation rate on this trapped layer is not examined. Since the beginning of the COVID-19 pandemic, numerous studies have been conducted using CFD tools to assess the risk of airborne infection and the effectiveness of risk mitigation strategies in various indoor environments, i.e., face masks [46-48] and ventilation, including public environments (i.e., restaurants [49], grocery stores [50, 62], classrooms [43, 51, 62]) and public transportation (passenger car [52], urban bus [53], car parking [54], aircraft cabin [23], elevator cabin [55, 62], subway [56]). In particular, using a Eulerian–Lagrangian simulation model, Ren et al. [57] found a local high deposition region in a prefabricated COVID-19 double-patient ward with a ceiling height of 2.60 m and suggested that outlet(s) should be installed inside the landing area of large particles and close to the polluted source(s). Most recently, using direct numerical simulation in a model room (3 m × 3 m) with a ceiling height of 3 m, Yang et al. [58] suggested that owing to the lock-up effects, increasing the ventilation rate does not further reduce the concentrations of pollutants above the critical ventilation rate and proposed an energy balance model to explain this phenomenon. However, the interaction between the occupant and thermal stratification interface is only moderately strong in this work since the ceiling is sufficiently high and the occupant is totally situated in the lower cool zone below the interface height.

Motivated by the abovementioned knowledge gap, in the present work, we conducted systematic studies of airborne transmission in a displacement ventilated low ceiling room (2.44 m) using Computational Fluid Dynamics (CFD) tools based on a Eulerian-Lagrangian approach, in particular the thermal lock-up effects on particle dispersion. The objective of this work is to 1) use CFD tools to systematically investigate the flow fields (i.e., thermal stratification) of displacement ventilation under various ventilation rates and how that affects the particle removal efficiency (risk levels) in a low ceiling room with occupants; 2) quantify the physical mechanism of interaction between ventilation flow and buoyancy flows under various ventilation rates and occupant locations and its influence on the critical ventilation rate where risk of infection is minimum (i.e., particle removal efficiency reaches its local maximum); 3) propose a new mitigation strategy to reduce the risk level of airborne infection, allowing us to be able to minimize the airborne transmission risk and keep low-ceiling rooms safe. The paper is structured as follows: Section 2 provides a description of the numerical methods, the geometry of the test model room, the exhaled aerosol particle injection model and case designs. Section 3.1 reports the results of flow fields and particle removal efficiency with varying ventilation rates and occupant locations. Section 3.2 reports the results of the influence of ventilation temperature on particle removal efficiency and the corresponding thermal interface height ($h_{ti}$) and proposes a ventilation strategy for ventilation design in terms of reducing the risk of airborne transmission. Section 4 provides a summary and discussion of the results.

## II. METHODS

### A. A test model room with low-ceiling and the spatial mesh

A three-dimensional (3D) numerical model of a small room with low-ceiling (2.44 m) was developed for the present study, i.e., a model of an elevator car is one type of low-ceiling room. The model room has overall dimensions of 2.03 (m) × 1.65 (m) × 2.44 (m) (L × W × H), as shown in detail in Fig. 1. The model is designed to yield the same dimensions as a typical elevator cabin with an asymptomatic occupant placed inside. The occupant is represented as a rectangular block with dimensions of 0.28 (m) × 0.16 (m) × 1.8 (m) (L × W × H), and stands on the floor facing the front wall (i.e., elevator door), as shown in Fig. 1. A previous study showed that the thermal interface height generated by the body plume of a real body shape agrees well with that of a simplified buoyance source in a model room with displacement ventilation [58], so we consider this simplification to be sufficient for the scope of the current study. The height of the mouth is 1.60 m above the floor, and the mouth opening is 3.14 cm$^2$, with a diameter of 10 mm. The total body surface area is 1.63 m$^2$. The simplified cuboid human body model is adopted from the models used in numerical simulations of the dispersion of human exhaled droplets under different ventilation methods in a classroom [59]. A structured mesh is used with near-wall refinement, as shown in Fig. 2. The mesh consists of $3.2 \times 10^6$ mesh cells with a maximum mesh size of ~15 mm and minimum of ~1 mm. Fine meshes of ~1 mm are used near the body surface, walls, ceiling, and floor to ensure the maximum $y^+ < 1$ during the simulation. The mesh near the mouth of the asymptomatic person is further refined to approximately 1 mm to capture the breathing jet dynamics. To verify that the computations are sufficiently independent from the mesh, the results obtained for one case using the current mesh for one occupant case with $3.2 \times 10^6$ were compared against those obtained using a much finer mesh with $3.6 \times 10^6$ mesh with a maximum mesh size of ~10 mm while keeping the minimum mesh size the same as presented in the Appendix. The differences observed do not suggest the need for additional refinement of the base mesh.

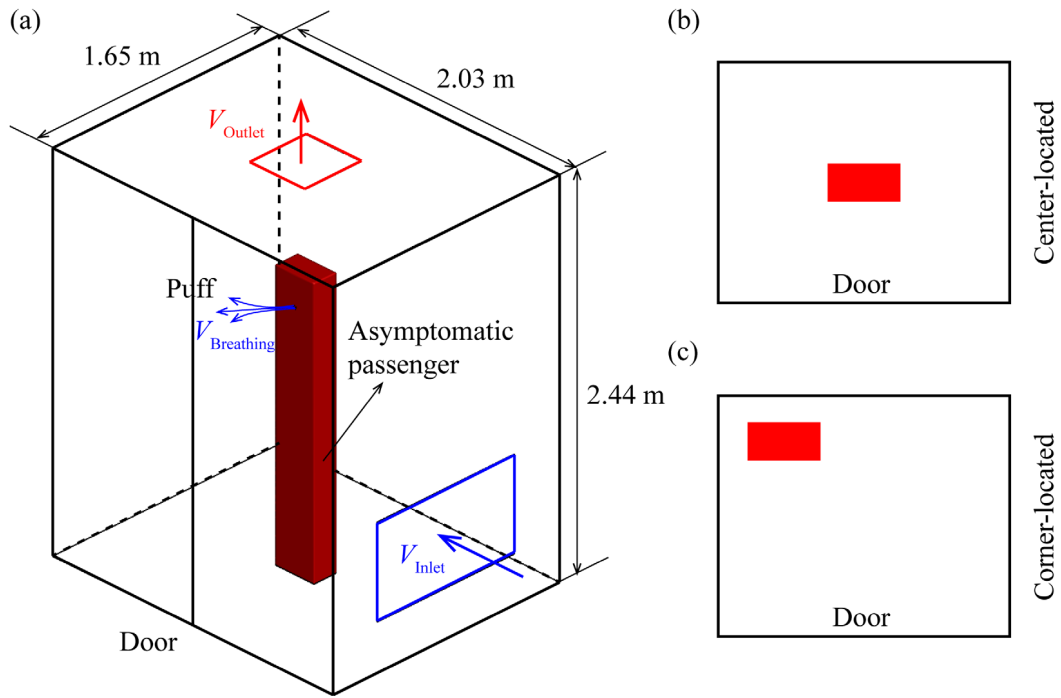

FIG. 1 Schematics of (a) the computational domain with one standing asymptomatic occupant inside a model room and the corresponding ventilation and respiratory flow settings, and top views showing (b) center-located occupant and (c) corner-located occupant settings. (Note that the red block represents the location and dimensions of the asymptomatic occupant).

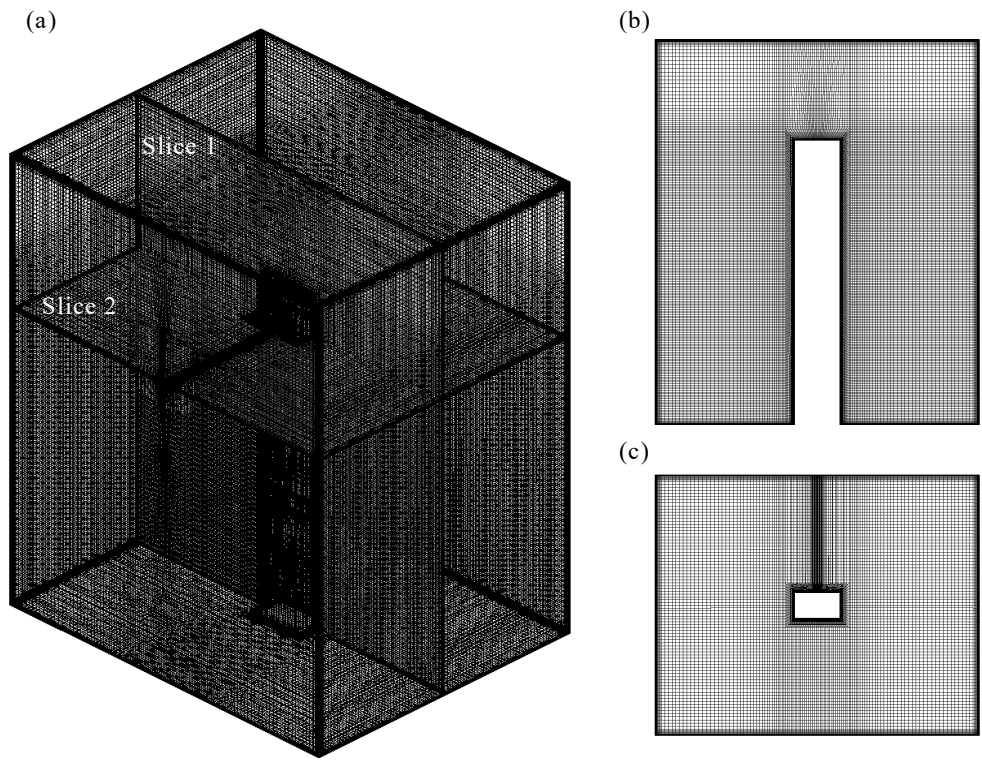

FIG. 2 A sample of computational meshes used for the present study including (a) 3D meshes and (b) the side view (slice 1) and (c) top view (slice 2) at the corresponding planes marked in (a) for the case corresponding to a center-located occupant.

## B. Numerical simulations

The present study uses the open source finite volume computational fluid dynamics code, OpenFOAM-6, *reactingParcelFoam*, to simulate the transient airflow and aerosol transport in the Eulerian-Lagrangian framework where the Lagrangian tracking calculation is coupled together with the flow solver. The transient airflow is calculated by solving the compressible Navier-Stokes equations, including mass conservation, momentum conservation and energy conservation equations of the carrier phase, along with the species equation using Eulerian descriptions. The transmission of respiratory aerosols containing the viruses is tracked by solving the advection equation of the discrete phase using the Lagrangian approach. One-way coupling is used to consider the interactions between aerosols and the carrier phase, i.e., the air flow is not affected by the motion of the aerosols, which is valid when the volume fraction is less than $10^{-6}$ in particle-laden turbulent flows [60] and is also adopted in a coughing jet study [61], where particle concentration is shown to be higher than that of normal breathing.

The governing equations of the carrier phase (airflow) are given by the following equations:

$$\frac{\partial \rho}{\partial t} + \nabla \cdot (\rho U) = \dot{S}_m \tag{1}$$

$$\frac{\partial \rho U}{\partial t} + \nabla \cdot (\rho U U) = -\nabla p + \nabla \cdot \tau + \rho g + \dot{S}_u \tag{2}$$

$$\frac{\partial \rho h}{\partial t} + \nabla \cdot (\rho U h) + \frac{\partial \rho K}{\partial t} \nabla \cdot (\rho K U) = \nabla(\alpha_{eff} \nabla h) + \rho U \cdot g + \frac{\partial p}{\partial t} + \dot{S}_h \tag{3}$$

where $\rho$ is the fluid density, U represents the velocity, $g$=9.81 m/s$^2$ is the gravity acceleration, $p$ is the pressure, $h = h_s + \sum_i c_i \Delta h_f^k$ ($h_s$=$c_p T$ is the sensible enthalpy, $c_i$ and $\Delta h_f^k$ are the molar fraction and the standard enthalpy of formation of species $i$, respectively) is the enthalpy, $K \equiv \|u\|^2/2$ is the kinetic energy per unit mass, and $\alpha_{eff} = \frac{\mu}{Pr} + \frac{\rho \nu_t}{Pr_t} = \frac{K_p}{\rho c_p} + \frac{\rho \nu_t}{Pr_t}$ ($K_p$ is the thermal conductivity, $c_p$ is the specific heat at constant pressure, Pr is the Prandtl number, Pr$_t$ is the turbulent Prandtl number, $\mu$ is the laminar dynamic viscosity and $\nu_t$ is the turbulent eddy viscosity) is the effective thermal diffusivity. In the above equations, $\dot{S}_m$, $\dot{S}_u$ and $\dot{S}_h$ are the mass, momentum and energy source terms generated by the aerosols, respectively. It is noted that the dynamics at the aerosol scale, such as break-up and evaporation, are not considered in this work, assuming aerosols have already reduced to the minimum size (mode of aerosol size is 1.7 μm) [62], and these aerosol-induced terms are zero during the simulation. In OpenFOAM, the pressure gradient and gravity force terms are rearranged numerically as $-\nabla p + \rho g = -\nabla p_{rgh} -$ $(g \cdot r)\nabla\rho$, where $p_{rgh} = p - \rho g \cdot r$ and $r$ is the position vector from the wall. $\tau$ is the stress tensor given by

$$\tau = \mu_{eff}\left(\nabla U + (\nabla U)^T - \frac{2}{3}\nabla(U)\right) \tag{4}$$

where $\mu_{eff} = \mu + \mu_t$ is the effective viscosity and $\mu_t = \rho \nu_t$. The turbulent viscosity, $\nu_t$, is modeled using $k$-$\omega$ Shear Stress Transport (SST) [63], which has been used in normal respiration as previously described [23, 25]. In the simulation, the relative humidity (RH) is 40% in the air. The carrier phase, which consists of two species (i.e., $H_2O$ and air), is modeled by their mass fraction $Y_k$ (gaseous) and is calculated from the equation of species in the flow fields:

$$\frac{\partial\rho Y_k}{\partial t} + \nabla \cdot (\rho Y_k u) = \nabla(\rho D_k \nabla Y_k) + \dot{\omega}_k \tag{5}$$

where $D_k$ is a diffusion coefficient of the $k$ species and $\dot{\omega}_k$ is a source term describing the generation of a species, and this term is zero assuming no species generation during the simulation. In the simulation, the Crank-Nicolson scheme with a bending coefficient of 0.9 was used for the time deviation term. The second-order upwind scheme is used for convective terms, and the Gauss-linear second order approach is used for the diffusion terms. The pressure-velocity coupling is solved by the Pressure-Implicit with Splitting of Operator (PISO) algorithm [64]. Discretized equations are solved with the geometric algebraic multigrid (GAMG) method in conjunction with the Gauss–Seidel solver. The minimum residuals for the convergence of pressure and velocity were $10^{-8}$ and $10^{-12}$, respectively. Virus-laden particles are injected through normal breathing activities. The particle size distribution falls in the range of $[0.5\ \mu m, 50\ \mu m]$ with a mean diameter of $1.7\ \mu m$ following the Weibull distribution [62]. The aerosol trajectories are solved via a Lagrangian approach. Aerosol particles are assumed to be spherical and pure liquid. Rotation and particle-particle interactions were ignored according to the experimental data that the particle number density is too low for collision [62]. The translational motion of each particle is governed by the force balance equation, i.e., the Maxey-Riley equation [65]. Aerosol velocity $u_{i,P}$ and position $x_{i,P}$ are obtained from the solution of the force balance equation given by:

$$\frac{d\vec{x}_{i,p}}{\partial t} = \vec{u}_{i,p} + \vec{u}_{i,p,t} \tag{6}$$

$$m_{i,p}\frac{d\vec{u}_{i,p}}{\partial t} = F_i^D + F_i^L + F_i^{BM} + F_i^G \tag{7}$$

where $i$ is the particle ID, $u_p$ is the particle velocity, $u_{p,t}$ is the stochastic velocity due to turbulence, $m_p$ is the particle mass, $F^D$ represents the drag force [66], $F^L$ is the Saffman's lift force [67], $F^G$ is the gravitational force, and $F^{BM}$ is the Brownian motion induced force [68]. Other forces, such as virtual mass and Basset forces, are normally not relevant because of the high liquid/air density ratio. The perturbation velocity $u_{p,t}$ is calculated based on the stochastic dispersion model of [69], where the fluctuation in direction $i$ is calculated as

$$u_{i,p,t} = \sigma\sqrt{\frac{2k}{3}} \tag{8}$$

with $\sigma \sim N(0,1)$ following the standard normal distribution and $k$ is the turbulence kinematic energy obtained from the simulation data. The heat transfer between aerosols and airflow is calculated with the Ranz-Marshall model [70], and the heat transfer coefficient is as follows:

$$C_v(Re_p) = 2 + 0.6Re_p^{0.5}Pr^{0.33} \tag{9}$$

where Pr is the Prandtl number of the gas. The drag force based on the rigid sphere assumption is calculated according to the following equation:

$$F^D = \frac{1}{8}\rho_g C_D \pi d_p^2 |\vec{u}_p - \vec{u}_f|(\vec{u}_f - \vec{u}_p)$$ (10)

where $d_p$ is the particle diameter, $\vec{u}_f$ is the carrier phase velocity and $C_D$ is the drag coefficient, which is determined by the following equation [66]:

$$C_D = \begin{cases} 0.424, & Re_p \geq 1000 \\ \frac{24.0}{Re_p}\left(1 + \frac{1}{6}Re_p^{\frac{2}{3}}\right), & Re_p < 1000 \end{cases}$$ (11)

where $Re_p = \frac{\rho|\vec{u}_f - \vec{u}_p|d_p}{\mu}$ is the particle Reynolds number. The Saffman's lift force is defined as follows [67],

$$F^L = m_{i,p}\frac{2K_s v^{\frac{1}{2}}d_{ij}}{\frac{\rho_p}{\rho_f}d_p(d_{lk}d_{kl})^{\frac{1}{4}}}(\vec{u}_f - \vec{u}_p)$$ (12)

where $K_s$=2.594 is the constant coefficient of Suffman's lift force and $d_{ij}$ is the deformation rate tensor defined as $d_{ij} = \frac{1}{2}(u_{i,j} + u_{j,i})$. The Brownian motion-induced force is modeled as a Gaussian white noise process given by [68],

$$F^{BL} = m_{i,p}G_i\sqrt{\frac{\pi S_0}{\Delta t}}$$ (13)

where $G_i$ are the zero-mean, unit variance independent Gaussian random numbers, $\Delta t$ is the time step used in the simulation, and

$$S_0 = \frac{216 v k_B T}{\pi^2 \rho d_p^5\left(\frac{\rho_p}{\rho_f}\right)^2}$$ (14)

where $k_B$=1.38×10[-23] J/K is the Stefan-Boltzmann constant. The amplitudes of the Brownian force components are evaluated at each time step. The last term is the gravity force, which is dependent on the density ratio between liquid and aerosol particles given by

$$F^G = m_{i,p}g\left(1 - \frac{\rho_f}{\rho_p}\right)$$ (15)

For particle-wall interactions, since the particles are small enough to stick on the surfaces, the stick boundary condition is used for the particles over all the solid surfaces. An escape boundary condition is employed for the outlet of the ventilation system. A reflection boundary condition for the inlet of the ventilation system and the mouth of the asymptomatic occupant.

## C. Study designs

Table 1 summarizes all the simulation cases presented in the current study, including (i) cases to study the effects of ventilation rate at the center (location 1, Case 1 ~ Case 8) and at the corner (location 2, Case 9 ~ Case 16) and (ii) cases to evaluate the effect of ventilation temperature on the particle removal efficiency of ventilation system at location 1 (Case T1 and Case T3) and at location 2 (Case T2 and Case T4). Specifically, for ventilation rate effect cases, the low-ceiling model room is equipped with a displacement ventilation system (DV), as shown in Fig. 1. The inlet of the ventilation system is located at the sidewalls, 0.6 meters from the floor, and in the center of the width direction and the outlet at the ceiling. The inlet and outlet dimensions of DV are $1.6 \times 0.8$ m$^2$ and $0.4 \times 0.4$ m$^2$, respectively. A pressure boundary condition is applied to the outlet, while a constant velocity boundary condition is used for the inlet. A no-slip boundary condition is applied to the solid walls. The temperature inside the model room is set as 20 °C, and adiabatic boundary conditions are used to model the room sidewalls, floor, and ceiling, which means that there is no heat transfer through the solid walls to or from the outside, and the occupant is the only heat source in the room. The surface of the manikin is set to 31 °C [71], and the respiratory flow is set to 33 °C [72]. For cases in the study of ventilation rate effects, the heating or cooling effect of the ventilation system is not considered, so the temperature from the inlet of the ventilation system is the same as that from room temperature. An asymptomatic occupant, referred to as the infector hereafter, stands at location 1 and at location 2. The infector is the only heat source driving a thermal buoyancy flow in the model room. The breathing activities are modeled by applying a time-varying injection profile at the infector's mouth to mimic human breathing, and the flow rate of the respiratory flow is 0.20 L/s based on experimental data [62] for all the simulation cases, and particles are injected at 44 particles per breathing cycle, as shown in Fig. 3. The transient simulations are conducted over a 3-minute duration for particle injection, representing an infector standing in the space for three minutes. Consequently, for each case, a total 3-minute simulation is conducted. It should be noted that to ensure the statistical convergence of particle trajectories, the particle number used is 10 times from normal breathing activities, and thus 440 particles per breathing cycle.

**TABLE 1.** A summary of the simulation cases and their corresponding ventilation and infector locations in the present study.

| Investigation | Case No. | Ventilation type | Source | Ambient temperature (°C) | Ventilation temperature (°C) | Ventilation rate ($Q_v$, ACH) | Capacity |
|---|---|---|---|---|---|---|---|
| | Case 1 | DV | Location 1 | 20 | 20 | 3 | 1 |
| | Case 2 | DV | Location 1 | 20 | 20 | 5 | 1 |
| | Case 3 | DV | Location 1 | 20 | 20 | 7 | 1 |
| | Case 4 | DV | Location 1 | 20 | 20 | 9 | 1 |
| | Case 5 | DV | Location 1 | 20 | 20 | 12 | 1 |
| | Case 6 | DV | Location 1 | 20 | 20 | 15 | 1 |
| | Case 7 | DV | Location 1 | 20 | 20 | 18 | 1 |
| Effect of | Case 8 | DV | Location 1 | 20 | 20 | 21 | 1 |
| ventilation rate | Case 9 | DV | Location 2 | 20 | 20 | 3 | 1 |
| | Case 10 | DV | Location 2 | 20 | 20 | 5 | 1 |
| | Case 11 | DV | Location 2 | 20 | 20 | 7 | 1 |
| | Case 12 | DV | Location 2 | 20 | 20 | 9 | 1 |
| | Case 13 | DV | Location 2 | 20 | 20 | 12 | 1 |
| | Case 14 | DV | Location 2 | 20 | 20 | 15 | 1 |
| | Case 15 | DV | Location 2 | 20 | 20 | 18 | 1 |
| | Case 16 | DV | Location 2 | 20 | 20 | 21 | 1 |
| Effect of | Case T1 | DV | Location 1 | 20 | 19 | 7 | 1 |
| ventilation | Case T2 | DV | Location 2 | 20 | 21 | 15 | 1 |
| temperature | Case T3 | DV | Location 1 | 20 | 19 | 7 | 1 |
| | Case T4 | DV | Location 2 | 20 | 21 | 15 | 1 |
| | | | | | | * DV-displacement ventilation | |

To study the effects of the ventilation rate, the ventilation is gradually increased to achieve an increase in the effective air change from 3 ACH to 21 ACH. Specifically, the range of the ventilation rate is simulated from 6.81 L/s to 47.68 L/s corresponding to ACH from 3 to 21. To evaluate the ventilation temperature effect (heating or cooling effect), four additional cases (Case T1 ~ Case T4) corresponding to two infector locations with lower and higher ventilation temperatures than the ambient temperature are included. Two infector locations (i.e., locations 1 and 2) and two ventilation temperature settings are simulated in a total of four cases.

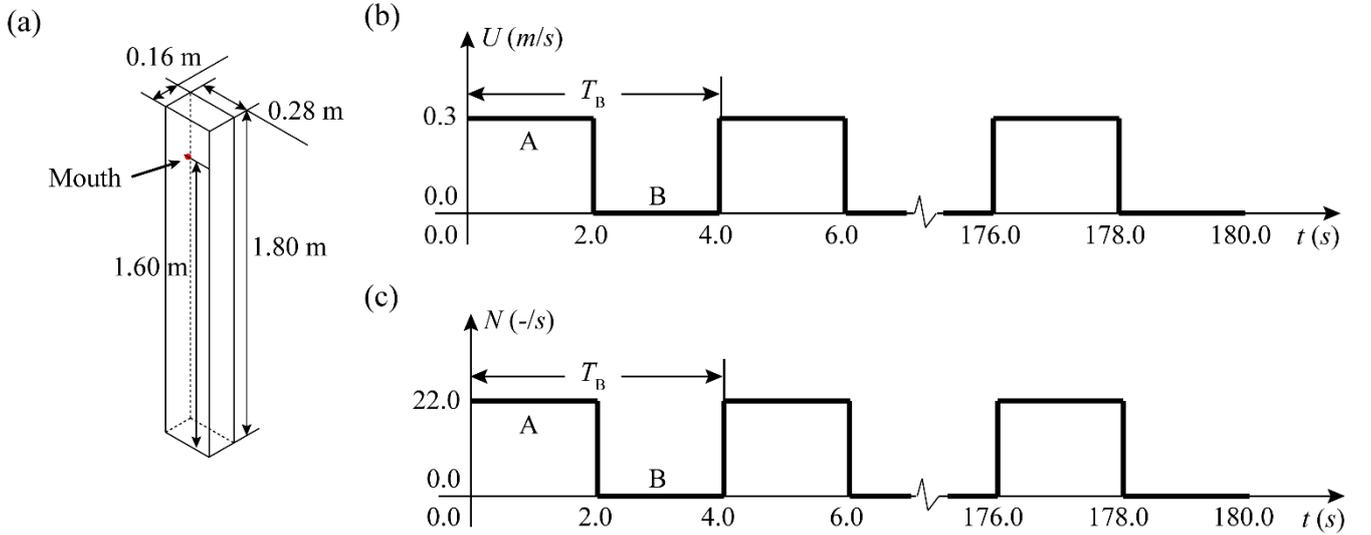

FIG. 3 (a) The body model for an asymptomatic occupant and the mouth height and the evolution of (b) the exhalation flow velocity and (c) the number of respiratory particles during the 3-minute simulation. A breathing cycle of a total of 4 s ($T_B$) with 2 s exhalation (A) and 2 s inhalation (B).

# III. RESULTS

## A. Effects of ventilation rate on the particle removal efficiency and flow fields

The particle removal efficiency and flow fields in a wide range of ventilation rates for various infector locations inside a low-ceiling test model room (i.e., elevator cabin) are examined, where the continuous phase is calculated using the Eulerian method and the particle trajectories are determined using Lagrangian tracking. Fig 4 presents the variation of particle removal efficiency ($\varepsilon_P$), i.e., the ratio of particles that are escaped from ventilation outlets to total particles injected inside the room, under different ventilation rates. This particle removal efficiency has been commonly used to evaluate the performance of ventilation for mitigating airborne transmission [17, 28, 73-75]. We find that for both center- and corner-located infector cases, with increasing $Q_v$ from 3 ACH to 21 ACH, $\varepsilon_P$ first increases sharply and then plateaus. Moreover, the center location shows a significantly higher plateaued value of particle removal efficiency ($\varepsilon_{P,c}$) and a lower plateaued value of ventilation rate compared to those of the corner location. Specifically, for the center-located infector, with increasing $Q_v$ from 3 ACH to 9 ACH, $\varepsilon_P$ increases sharply from 45.0% to 72.2% and then plateaus at approximately $Q_v$=9 ACH. With further increasing $Q_v$ from 9 ACH to 21 ACH, $\varepsilon_P$ keeps increasing but with a much lower rate and rises from 72.2% up to approximately 80.0%. For the corner-located infector, with increasing $Q_v$ from 3 ACH to 12 ACH, $\varepsilon_P$ increases from 7.8% to 46.6% and then plateaus at approximately 12 ACH. With further increasing $Q_v$ from 12 ACH to 21 ACH, $\varepsilon_P$ remains almost at a constant value different from that at the center-located infector and rises slightly from 46.6% to 53.0%. It is worth noting that the critical ventilation rate ($Q_c$), i.e., the ventilation rate where the particle removal efficiency starts to plateau, as indicated by the black shadow region for the center-located infector (approximately 72.2% at $Q_v$=9 ACH), is lower than that for the corner-located infector, as shown by the red shadow region (approximately 46.6% at $Q_v$=12 ACH). Correspondingly, $Q_c$ at the corner-located infector is larger than that at the center-located infector. There exists a shift of $Q_c$ from $Q_v$=9 ACH at the center-located infector to $Q_v$=12 ACH at the corner-located infector. These trends observed in Fig. 4 in terms of $\varepsilon_p$ with increasing $Q_v$ and the corresponding $Q_c$ will be explained later when we examine the flow fields in detail. To further investigate the particle fates under different ventilation rates and infector locations, the variation of particle deposition on walls (including floor, ceiling, and all four sidewalls) under different ventilation rates is examined. It is worth noting that the maximum particle deposition percentage is less than 2.0% for all ventilation rates and infector locations, indicating that the majority of particles (microns in size) in the current low-ceiling room either remain suspended or are removed out from the room by the ventilation system after 3 minutes of simulation. Moreover, with increasing $Q_v$ from 3 ACH to 21 ACH, the percentage of particle deposition on the wall increases slightly, i.e., from 0.5% to 1.8% at the center-located infector and from 0.3% to 1.5% at the corner-located infector, respectively, due to enhanced turbulent wall flux associated with increasing wall shear stress/friction velocity [76].

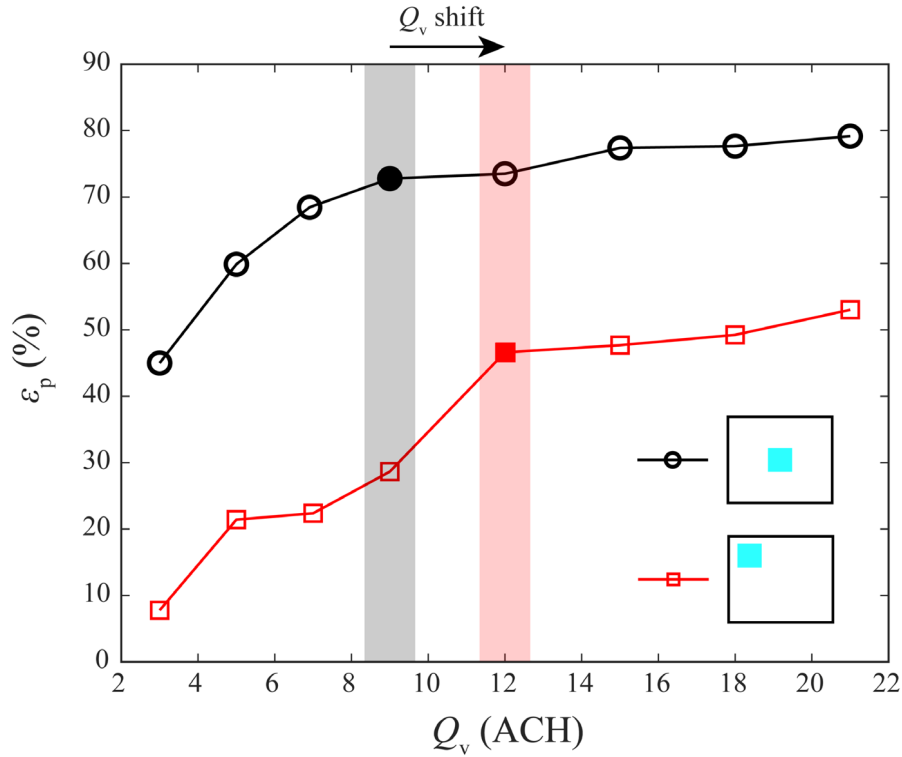

FIG. 4 The variation of particle removal efficiency ($\varepsilon_p$) under different ventilation rates ($Q_v$) for center- and corner-located infectors.

To elucidate the physical mechanism underlying the effects of $Q_v$ on $\varepsilon_p$ (i.e., $\varepsilon_{p,c}$), we examine the particle dispersion and flow fields for several cases (case 1, case 3, case 4, case 5, case 6, case 8, case 9, case 11, case 12, case 13, case 14, and case 16) where ventilation rates below, near and above $Q_c$ are studied. In Fig. 5, the particle dispersion inside the model room (top row) and the corresponding mean temperature (middle row) and horizontal velocity ($U_x$, bottom row) distributions on the $x$-$y$ plane crossing the middle of the infector are presented for the center-located infector for $Q_v$=3 ACH, 7 ACH, 9 ACH, 12 ACH, 15 ACH and 21 ACH. The mean flow fields are determined as the 3-minute time-averaged values. The upper warm layer is represented by the blue temperature isosurface of $T$=20.7 °C located at the infector mouth height at $Q_c$, which is 0.7 °C larger than the ambient room temperature, i.e., $T$=20.0 °C, and 10.3 °C and 12.3 °C less than the infector body and respiratory flow temperature, respectively. Here, the height of the temperature isosurface of $T$=20.7 °C is marked by $h_{T,20.7}$, and the height of the infector mouth is marked by $h_{im}$. Generally, with increasing $Q_v$, the upper warm layer moves toward the room ceiling along with increasing thermal stratified field, which is driven by the enhanced ventilation flows from the ventilation inlet, and the corresponding number of suspended particles decreases, as illustrated in Fig. 5 (*b*) ~ (*g*). Specifically, below $Q_c$=9 ACH, with increasing $Q_v$ from 3 ACH to 9 ACH, the upper warm layer as illustrated by $h_{T,20.7}$ is lifted upward, resulting in the volume of the upper warm layer shrinking and the particles being pushed out of the model room under the confinement effects of the room ceiling. Furthermore, the temperature distribution (middle row) on the *x-y* plane crossing the middle of the infector in the model room clearly shows the features of thermal stratification. It is worth noting that in the current low-ceiling room setting, $h_{T,20.7}<h_{im}$, the thermal interface between the lower cool layer and upper warm layer is lower than the infector height, and the infector mouth (i.e., virus-laden aerosol emission source) is located entirely inside the upper warm layer. Such a configuration leads to complex interactions between infector exhaled flows, ventilation flows and thermal plumes generated by the infector, which is different from the results in Yang et al. [58] for high ceiling rooms where the lower clean zone is above the infector and the infector mouth is located entirely inside the lower clean zone.

In Fig. 5, we find a close correlation between the variation of the thermal stratification field with $Q_v$ (i.e., $h_{T,20.7}$) and the corresponding $\varepsilon_p$, thus the risk of airborne infection. Clearly, with an increasing upper warm layer toward the room ceiling, the local thermal microenvironments (i.e., temperature and temperature gradient) near the infector mouth are significantly altered, leading to an increasing local temperature gradient and decreasing temperature difference between the infector body and local room temperature (i.e., $\Delta T_{ib\_tr}$) at the infector mouth height where virus-laden aerosol particles are injected, thus changing the particle dispersion. To quantitatively illustrate the variations of the temperature gradient, in Fig. 6, the variation of the local ($\partial T_m/\partial H$, $H$=1.60 m) and mean temperature gradients ($\Delta T_{m\_mean}/\Delta H$, $H$=1.50 ~ 1.70 m) in the breathing zone near the infector mouth is plotted for various ventilation rates. The breathing zone is defined as the region with $H$=1.40~1.80 m, which is the typical height range covering the location of the infector mouth from adults to children. In Fig. 6, we find that with increasing $Q_v$, both the local and mean temperature gradients at mouth height first increase, then plateau at approximately 9 ACH, and finally decrease sharply. Remarkably, the plateaued values of $\partial T_m/\partial H$ and $\Delta T_{m\_mean}/\Delta H$ settle at $Q_c$, where the maximum particle removal efficiency occurs, indicating that the temperature gradient near the infector mouth significantly influences the particle removal efficiency. In particular, we find that at $Q_c$, where the particle removal efficiency plateaus, $h_{T,20.7}$ arrives at the infector mouth, as shown by the blue isosurface in Fig. 5, along with that the temperature gradient at mouth height is approximately 1.68 °C/m, the mean temperature gradient in the breathing zone is 1.52 °C/m, and $\Delta T_{ib\_tr}$ at $h_{im}$ is approximately 10.4 °C.

Above $Q_c$, with increasing $Q_v$ from 9 ACH to 21 ACH, the upper warm layer moves higher toward the room ceiling, as shown in Fig. 5, resulting in the further enlargement of the lower cool air layer and volume shrinkage of the upper warm air layer under the confinement of the room ceiling. Consequently, the upper warm air layer significantly prevents the aerosol particles from being vented out, that is, the lockup effect, and the particles suspended inside the warm air layer decrease. In this configuration, $h_{T,20.7}>h_{im}$, and the infector mouth is located inside the lower cool layer. Consequently, the room temperature at $h_{im}$ decreases (i.e., < 20.7 °C), and at $h_{im}$, $\Delta T_{ib\_tr}>10.4$ °C, as presented in the outlines of the mean temperature distribution in Fig. 5. Furthermore, both $\partial T_m/\partial H$ and $\Delta T_{m\_mean}/\Delta H$ decrease with increasing $Q_v$, as illustrated in Fig. 6. Additionally, the flows around the infector body (i.e., body side, above head) change significantly as $Q_v$ increases, which traps particles there and prevents them from being ventilated out. For example, the flow separation region on the ventilation side of the infector body, which initiates at approximately $Q_v$=7 ACH, is directly influenced by the ventilated flows, and the flow separation region on the other side of the infector body is enhanced in both strength and size, as shown in the contours of mean horizontal velocity ($U_x$) in the bottom row of Fig. 5. Compared with that below $Q_c$ of 9 ACH, the local flow recirculation occurs above the infector head and is enhanced with increasing $Q_v$ shown in both the mean temperature and mean horizontal velocity fields ($U_x$), as indicated by the red dashed ellipses in Fig. 5.

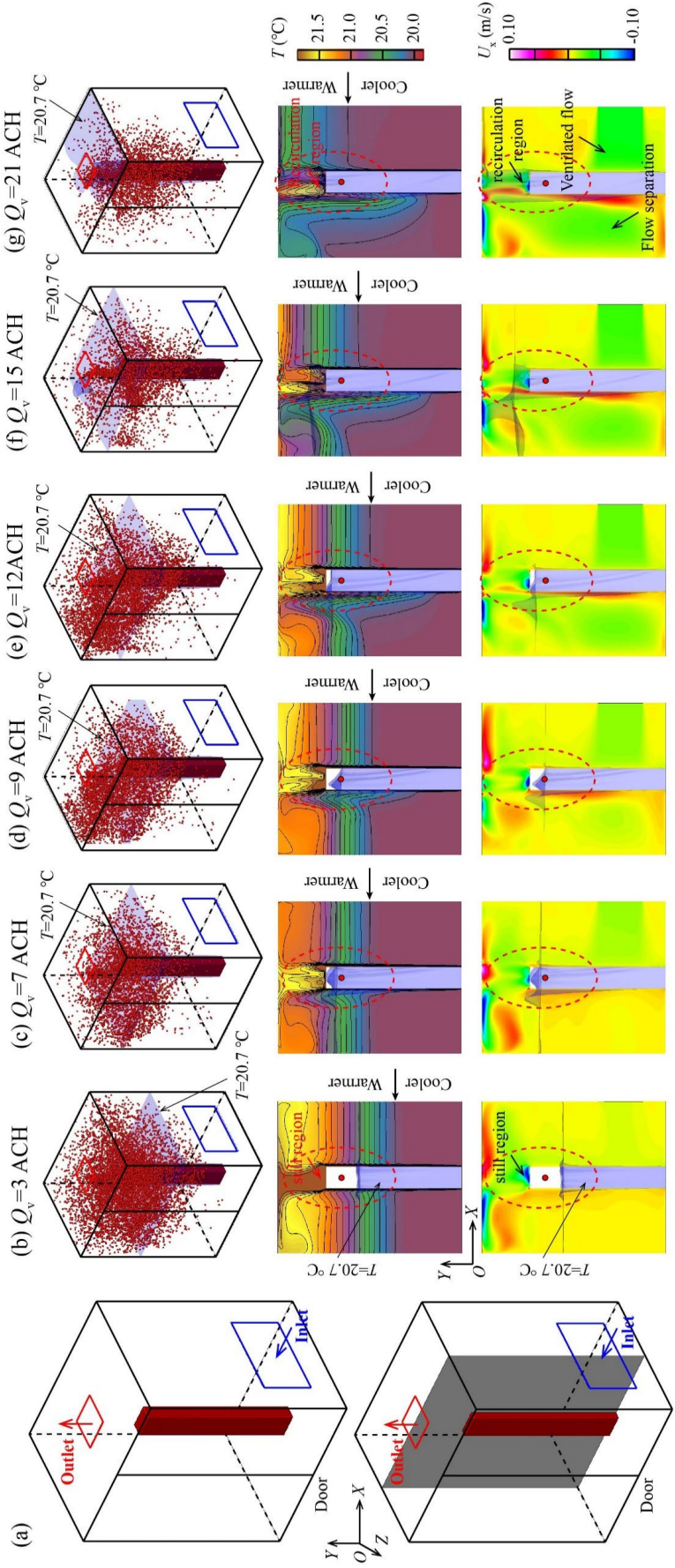

FIG. 5 Influence of ventilation rate on particle dispersion and flow fields for center-located infector as illustrated in the schematic of cases of the model room where arrows indicate the ventilation inlet (blue) and outlet (red) at the upside in ($a$), and ($b$–$g$) particle dispersion at the underside in ($a$). Red dashed ellipse indicates the microenvironment near infector mouth (e.g., aerosol emission source) region. blue region, $T$=20.7 °C), temperature (middle row), and horizontal velocity ($U_x$) of air flows (bottom row), respectively, on the $x$-$y$ plane crossing the middle of infector in the model room as illustrated at the underside in ($a$) for ($b$) $Q_v$=3 ACH, ($c$) $Q_v$=7 ACH, ($d$) $Q_v$=9 ACH, ($e$) $Q_v$=12 ACH, ($f$) $Q_v$=15 ACH, and ($g$) $Q_v$=21 where arrows indicate the ventilation inlet (blue) and outlet (red) at the upside in ($a$), and ($b$–$g$) particle dispersion along with the temperature isosurface (top row, ACH. Red dashed ellipse indicates the microenvironment near infector mouth (e.g., aerosol emission source) region.

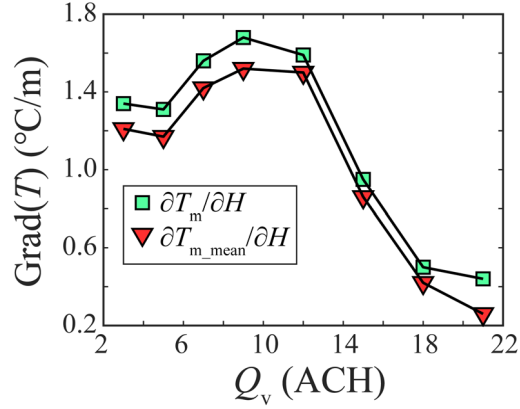

FIG. 6 The variation of local ($\partial T_m/\partial H$) and mean temperature ($\Delta T_{m\_mean}/\Delta H$) gradients under different ventilation rates for center-located infectors.

Consequently, below $Q_c$, the thermal stratification interface is relatively lower where the infector mouth is inside the high temperature layer (i.e., > 20.7 °C). With increasing $Q_v$, $h_{T,20.7}$ increases quickly, and exhaled aerosol particles are pushed upward toward the room ceiling and then ventilated out of the room, leading to a steep rise in $\varepsilon_p$. Above $Q_c$, the thermal stratification interface rises above the respiratory region where aerosols are injected, and the infector mouth is inside the low temperature layer (i.e., < 20.7 °C), causing the aerosol particle lockup effect. With increasing $Q_v$, the further rise of the thermal stratification interface is confined by the room ceiling, and part of the particles could be trapped inside the lower cool air layer, i.e., the lockup effect. Additionally, flow separation where a large recirculation region occurs near the infector body and above the body head is enhanced, and the flow separation region increases in size where particles can be trapped. Thus, the increase rate of $\varepsilon_p$ with increasing $Q_v$ above $Q_c$ is slower than that below $Q_c$.

To further elucidate the physical mechanisms involved in the connection between the variation in thermal stratification fields and $\varepsilon_p$, the flow fields, including mean temperature (top row), mean horizontal velocity ($U_z$, $U_x$, middle row), and mean vertical velocity ($U_y$, bottom row), on the $z$-$y$ plane crossing the body centerline are examined in Fig. 7 for the ventilation conditions corresponding to those in Fig. 5. The changes in complex interactions between body thermal plumes, respiratory flows and ventilation flows with increasing $Q_v$ can be well illustrated on the $y$-$z$ plane. Below $Q_c$ of 9 ACH, at $Q_v$=3 ACH and $Q_v$=7 ACH, the thermal stratification is clearly shown on the mean temperature contour (top row). Two streams of vertical velocities generated from the front and back of the infector, as indicated by the black arrows in the mean vertical velocity contours, are observed (middle row). At lower $Q_v$, the vertical velocity ($U_y$) in the vicinity of the body is stronger in the front than in the back, and this difference is significantly reduced when $Q_v$ rises above $Q_c$ (i.e., 9 ACH). There exists a low velocity region (i.e., flow separation and recirculation) above the infector head, as indicated by the red dashed ellipses in Fig. 7. At the same time, from $Q_v$=3 ACH to $Q_v$=7 ACH, the flow fields in the upper warm layer change substantially. For example, $h_{T,20.7}$ increases, and the strength of the stream of mean vertical velocity generated from the infector's mouth becomes stronger, while the other one generated from the infector back stays relatively weak. However, with further increasing $Q_v$ from 7 ACH to 9 ACH, $h_{T,20.7}$ decreases slightly, and the low velocity region above the body head enlarges. The strength of the streams of vertical velocity contour generated from both the infector front and back becomes stronger. Above $Q_c$ of 9 ACH, the same as that in Fig. 5, the thermal interface (i.e., $h_{T,20.7}$) keeps rising where the lock up effect occurs, and the local flows above the body head become complex and recirculate, leading to a stronger flow separation around the body. Different from that below $Q_c$, the vertical velocities of the two streams from both the back and front of the infector body are relatively weak. In summary, we find that $\varepsilon_p$ is largely associated with the strength of rising buoyancy flows near the body and that such flow is greatly suppressed due to the enhanced ventilated flows as $Q_v$ increases above a critical value $Q_c$.

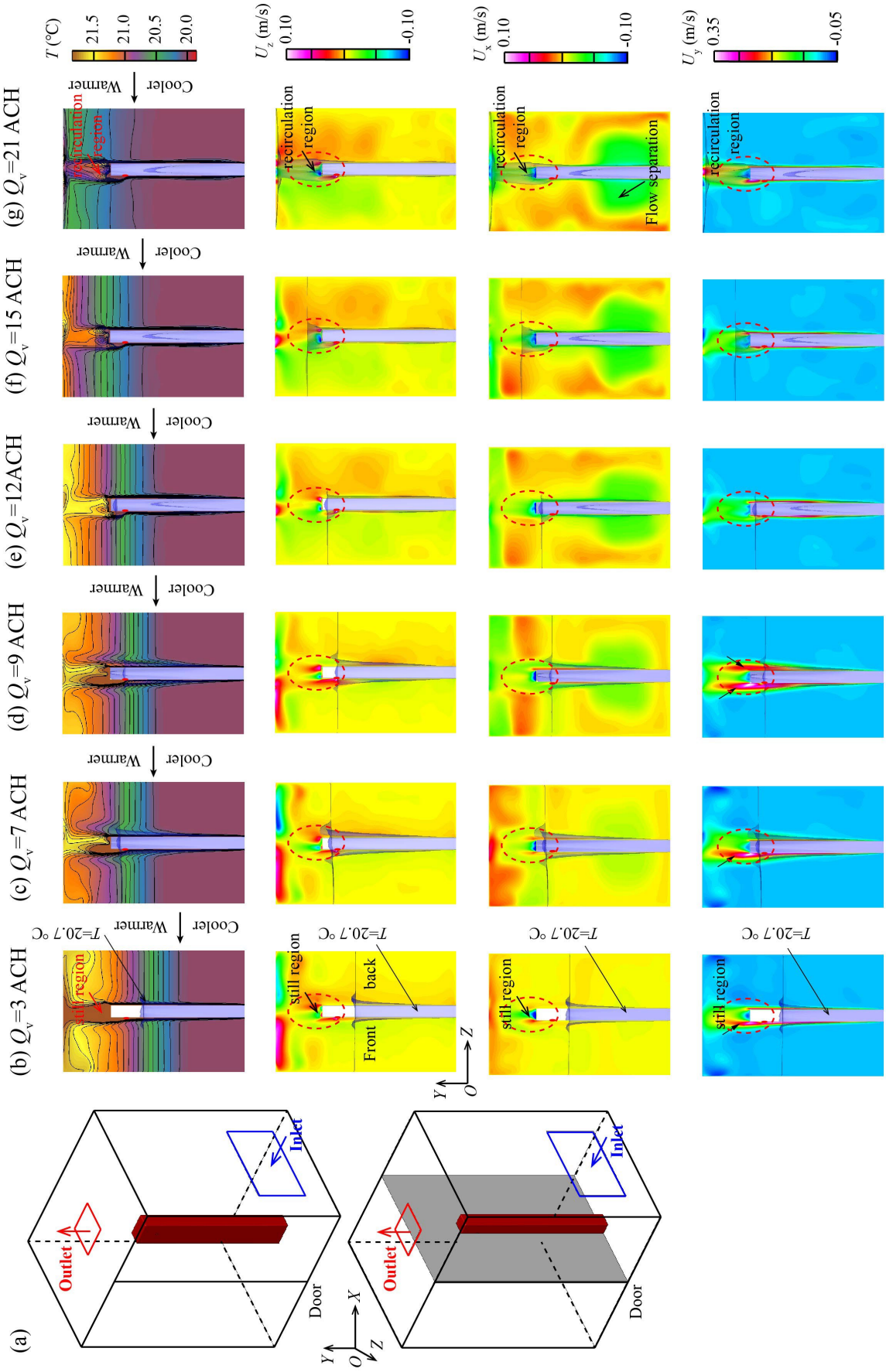

FIG. 7 The variation of mean temperature (top row), mean horizontal velocity ($U_x$ and $U_z$, middle row), and mean vertical velocity ($U_y$, bottom row) along with the temperature isosurface of $T=20.7\ ^{\circ}C$ on the $y$-$z$ plane crossing the middle of infector in the model room as illustrated in the underside of ($a$) for ($b$) $Q_v=3$ ACH, ($c$) $Q_v=7$ ACH, ($d$) $Q_v=9$ ACH, ($e$) $Q_v=12$ ACH, ($f$) $Q_v=15$ ACH and ($g$) $Q_v=21$ ACH for the center-located infector as illustrated at the upside of ($a$).

Since the infector location is important for the airborne transmission of infection risk in indoor environments, the particle dispersion and flow fields under the corner-located infector with the same ventilation rate range as the center-located infector conditions are examined in detail below. In Fig. 8, the particle dispersion inside the model room and the corresponding mean temperature distributions on the $x$-$y$ plane crossing the center of the infector body, as illustrated in Fig. 8, are examined for the corner-located infector at $Q_v$=3 ACH, 7 ACH, 9 ACH, 12 ACH, 15 ACH and 21 ACH. The stratified temperature field shows a lower cool layer and an upper warm layer, as indicated by the blue temperature isosurface of $T$=20.7 °C, which are similar to those for the center-located infector case in Fig. 5. With increasing $Q_v$, both the particle dispersion and temperature distribution change correspondingly. Specifically, as $Q_v$ increases from 3 to $Q_c$ of 12 ACH, $h_{si}$ gradually rises and reaches $h_{im}$, where $\varepsilon_p$ starts to plateau. This trend is similar to that in the center-located infector case (Figs. 5 and 7), but the $Q_c$ for $\varepsilon_p$ to plateau is higher. Furthermore, the temperature inside the upper warm layer gradually decreases with increasing $Q_v$. Above $Q_c$ of $Q_v$=12 ACH, $h_{T,20.7}$ > $h_{im}$, with further increasing ventilation rate from $Q_v$=12 ACH to $Q_v$=21 ACH, $h_{T,20.7}$ keeps increasing and the temperature in the up warmer layer decreases. Corresponding to $\varepsilon_p$ in Fig. 4, the variation of $\varepsilon_p$ with $Q_v$ is in agreement with that of $h_{T,20.7}$, and $\varepsilon_p$ starts to plateau at $Q_v$=12 ACH, where the temperature isosurface $T$=20.7 °C reaches $h_{im}$ and $\Delta T_{ib\_lr}$≈10.35 °C, which is also observed for center-located infector conditions. Consequently, there exists a close relationship between $\varepsilon_p$ and $\Delta T_{ib\_lr}$. Compared with that for the center-located infector conditions in Fig. 5, at the same ventilation rate, $h_{T,20.7}$ at the corner-located infector is lower than that at the center-located infector, and correspondingly, $\varepsilon_p$ is lower for the corner-located infector where the aerosol transmission path from the infector mouth to the ventilation outlet at room ceiling is longer. Furthermore, there is a shift of $Q_c$ from 9 ACH to 12 ACH at plateaued particle removal efficiency, which could be caused by the lower $h_{T,20.7}$ at corner-located infector conditions where more ventilation flows are required for the thermal stratification fields (i.e., $h_{T,20.7}$) to reach the same height. Because the infector is not located just in front of the ventilation inlet, the enhancement of flow separation near the infector body for the corner-located infector is weaker than that for the center-located infector.

FIG. 8 Influence of ventilation rate on particle dispersion and flow fields for corner-located infector as illustrated in the schematic of cases of the model room where arrows indicate the ventilation inlet and outlet at the upside in (a), and (b-g) particle dispersion along with the temperature isosurface (top row, blue region, $T$=20.7 °C), mean temperature (middle row), and horizontal velocity of air flows ($U_x$, bottom row), respectively, on the x-y plane crossing the middle of infector in the model room as illustrated at the underside in (a) for (b) $Q_v$=3 ACH, (c) $Q_v$=7 ACH, (d) $Q_v$=9 ACH, (e) $Q_v$=12 ACH, (f) $Q_v$=15 ACH, and (g) $Q_v$=21 ACH.

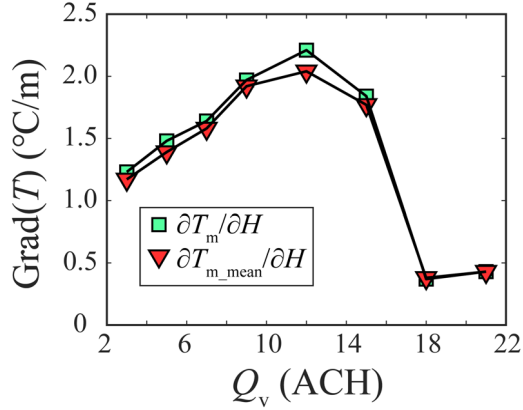

FIG. 9 The variation of local ($\partial T_m/\partial H$) and mean temperature gradients ($\Delta T_{m\_mean}/\Delta H$) under different ventilation rates for center-located infectors.

To further elucidate the physical mechanisms involved in the variations of thermal stratification fields (i.e., $h_{T,20,7}$) and $\varepsilon_p$ for the corner-located infector responsible for the observations in Fig. 4, Fig. 10 presents the flow fields, including mean temperature (top row), mean horizontal velocity ($U_x$, and $U_z$, middle row) and mean vertical velocity (bottom row), on the $y$-$z$ plane crossing the center of the infector body for the ventilation conditions corresponding to those in Fig. 8. As shown in Fig. 10, the temperature contour is characterized by strong stratification with a lower cool layer and upper warm layer, and convective buoyancy flows generated by the infector body and its interactions with the thermal stratification fields can be observed clearly. At a ventilation rate of $Q_v$=12 ACH for $\varepsilon_{p,c}$, $h_{T,20.7} \approx h_{im}$, a relatively high velocity region, as indicated by the red dashed ellipses, is observed on the mean streamwise velocity contour. Two streams of velocity generated from the infector front and back are marked by black arrows on the vertical velocity contour. It should be noted that the flow fields, especially the mean streamwise velocity and mean vertical velocity, for the corner-located infector are different from those for the center-located infector owing to the confinement effects of the room sidewalls and present nonsymmetric features. For example, the streamwise velocity above the infector head at $Q_v$=12 ACH for a corner-located infector is higher than that for a center-located infector, which is a dead still region with low velocity. The velocity stream generated from the body front is weaker than that from the body back for a corner-located infector, while for a center-located infector, the velocity stream generated from the body front is stronger than that from the body back. Below the $Q_c$ of $Q_v$=12 ACH, with increasing $Q_v$ from 3 ACH to 12 ACH, $h_{T,20.7}$ increases and the velocity intensity at the high streamwise velocity region above the body head decreases, while the stream of vertical velocity generated from the body front decreases and that from the body back increases. Above $Q_c$ of $Q_v$=12 ACH, with further increasing $Q_v$ from 12 ACH to 21 ACH, $h_{T,20.7}$ keeps increasing toward the room ceiling, the stream of vertical velocity generated from body back is further enhanced and that from the body front becomes weaker, which is caused by the infector location showing different interactions between the body plumes, ventilation flows and exhaled flows for a corner-located infector.

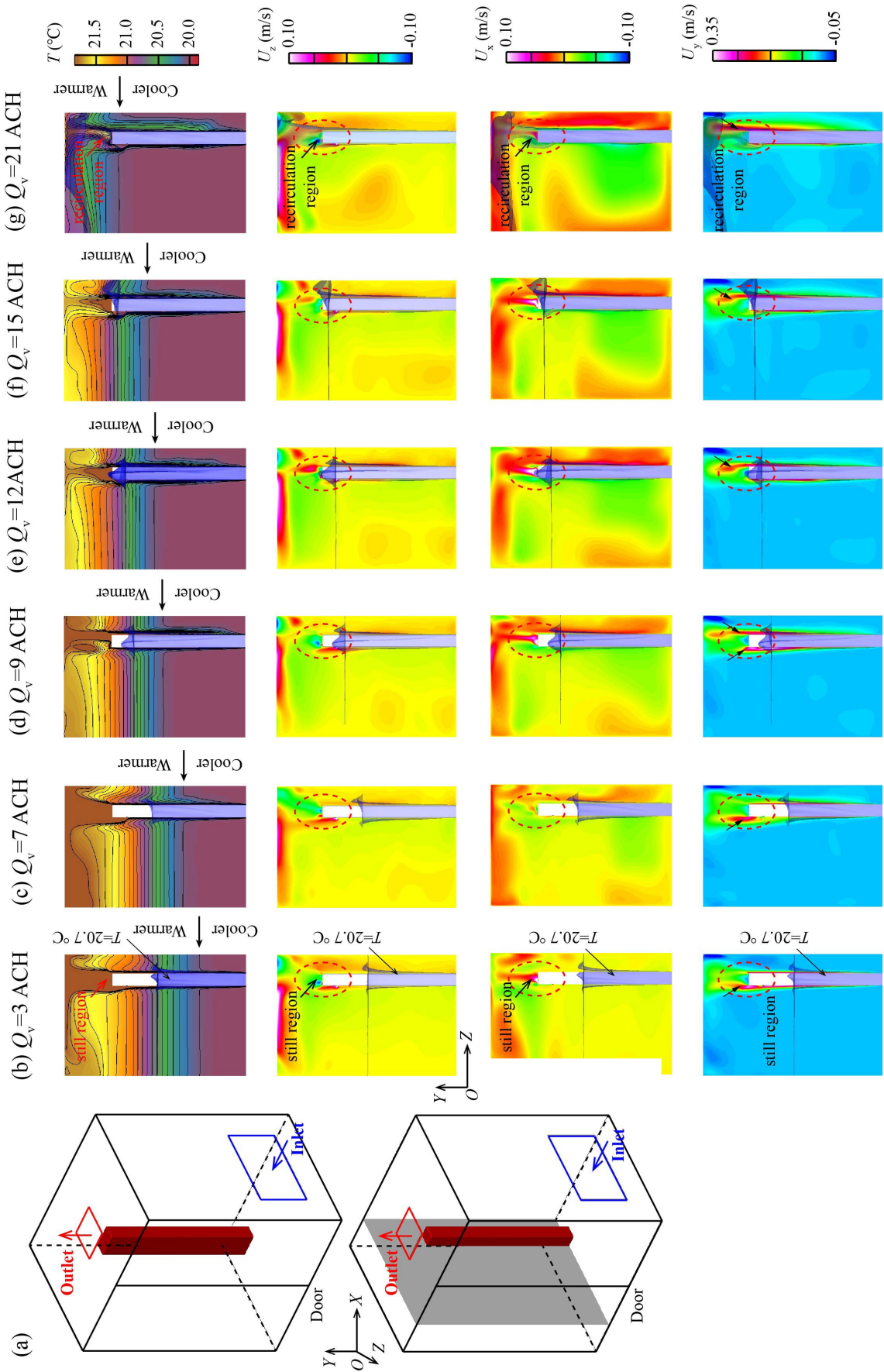

FIG. 10 The variation of mean temperature (top row), mean horizontal velocity ($U_x$, and $U_z$, middle row) and mean vertical velocity ($U_y$, bottom row) along with the temperature isosurface of $T=20.7\,°C$ on the $y$-$z$ plane crossing the middle of infector in the model room as illustrated at the underside of ($a$) for ($b$) $Q_v=3$ ACH, ($c$) $Q_v=7$ ACH, ($d$) $Q_v=9$ ACH, ($e$) $Q_v=12$ ACH, ($f$) $Q_v=15$ ACH and ($g$) $Q_v=21$ ACH for the corner-located infector as illustrated at the upside of ($a$).

From the above, we find that the thermal stratification fields (i.e., $h_{T,20.7}$) are crucial for aerosol particle dispersion and particle removal efficiency, which vary according to the infector locations. To follow up, we investigate the physics governing the variation of thermal interface height (i.e., $h_{ti}$) with ventilation rate ($Q_v$) under different infector locations. Here, $h_{ti}$ is calculated based on the mean temperature gradient profile obtained by averaging the computational cells at the same vertical height from the 3-minute average flow field, where the thermal interface height is located at the first inflection point from the ground and the temperature increase with respect to the vertical height is approximately 0.2 °C higher than the room temperature (i.e., 20.0 °C). In Fig. 11 (a), the variation of $h_{ti}$ with $\widetilde{Q_v}$ is presented, and in Fig. 11 (b) and Fig. 11 (c), the corresponding mean temperature and particle number profiles under both center- and corner-located infector cases are further examined to illustrate the lockup effect. To be consistent with the literature [36, 37, 38, 58] and facilitate comparison, we introduce the ventilation rate per person $\widetilde{Q_v}$. In Fig. 11, we find that at low ventilation rate conditions, the variation of $h_{ti}$ with $\widetilde{Q_v}$ is comparable with the classical $h_{ti} \sim \widetilde{Q_v}^{3/5}$ formula [36, 37, 38]. For the center-located infector case, when $\widetilde{Q_v} < 20.40$ L/s, $h_{ti}$ increases from $\widetilde{Q_v} = 7.00$ L/s (i.e., 3 ACH) to 20.40 L/s (i.e., 9 ACH), and the rising trend agrees with the $h_{ti} \sim \widetilde{Q_v}^{3/5}$ formula shown by the black dashed line in Fig. 11 (a), which is also confirmed by Yang et al. [58] in the high ceiling room where the infector is totally located in the low cooler thermal layer below $h_{ti}$. At the same time, the temperature along the vertical direction decreases, and the corresponding particle number reduces quickly. This finding indicates that for low ventilation rate conditions, the shape of the body and the interaction between the thermal interface and body plume have little effect on thermal stratification fields and thus $h_{ti}$. However, when $\widetilde{Q_v} > 20.40$ L/s, $h_{ti}$ begins to deviate from the $h_{ti} \sim \widetilde{Q_v}^{3/5}$ line and plateau. Meanwhile, both the temperature and the corresponding particle number keep decreasing slowly, which is in agreement with the variation of $h_{ti}$ with increasing $\widetilde{Q_v}$. This result suggests that the simplified heat source assumption used in $h_{ti} \sim \widetilde{Q_v}^{3/5}$ does not hold at high ventilation rate conditions, under which the interaction between the body plume and the stratified thermal layer cannot be ignored. Similar trends are also found for the corner-located infector. For the corner-located infector case, the variation of $h_{ti}$ with $\widetilde{Q_v}$ shows trends similar to those for the center-located infector except for a shift in $\widetilde{Q_c}$ from 20.40 L/s (9 ACH) to 27.20 L/s (12 ACH), showing that $\widetilde{Q_c}$ significantly depends on the infector location. In addition, we notice that above $\widetilde{Q_c}$, $h_{ti}$ for the center- and corner-located infector keeps decreasing slightly, and the lockup effects are evident for both center- and corner-located infector cases at high ventilation rates where the infector mouth is located inside the lower cool air layer with low temperature and large temperature gradient exists. The shift of $\widetilde{Q_c}$ due to the infector location will be further analyzed in detail by analyzing the turbulent kinematic energy fields as follows.

To elucidate the mechanism that leads to the shift of $Q_c$ when the location of the infector changes from center to corner location, the spatial variation of velocity fluctuations in terms of variance (Var) is presented to show the flow unsteadiness for both center- and corner-located infector in Fig. 12. Generally, the high intensity region of Var($U$) is located around the infector body, especially above the infector head, which is mainly driven by the convective buoyancy flows (i.e., thermal plume) generated by the infector body. The infector locations also significantly influence the spatial distributions of Var($U$). Specifically, for the center-located infector, below $Q_c$, the plume initially is highly unsteady and oscillates strongly around its equilibrium position along with a large local velocity fluctuation, Var($U$), especially in the regions where the plumes rise from the infector body (indicated by the magenta arrows), which leads to a strong dispersion of particles. At this stage, the plume is not entirely turbulent but experiences high unsteadiness. With increasing $Q_v$, Var($U$) keeps increasing, showing that the flow field is highly unsteady.

Above $Q_c$, the plume becomes stabilized, large oscillations disappear, and the plume may transition to a developed turbulent state. Consequently, the high Var($U$) region shrinks with further increasing $Q_v$ and gradually becomes confined above the body head, which is associated with the reduction in the large-scale oscillation of the body thermal plume, resulting in limited particle dispersion and thus preventing particles from being extracted out through the ventilation outlet. The shift of $Q_c$ with location is associated with the delay in the transition of the body plume from a large oscillation state to a developed turbulent state. This earlier transition of the body thermal plume from the unsteady state to the developed turbulent state observed in the center-located infector case may be caused by the stronger interaction between ventilation flow and the body thermal plume. For the corner-located infector, such an interaction is weaker since the infector body is located farther away from the ventilation outlet, which leads to the delay in the transition and thus the shift of $Q_c$. It should be noted that this $\widetilde{Q_c}$ could also be influenced by the number of occupants, vent size and positions, room shape and size, etc., as pointed out in Yang et al. [58].

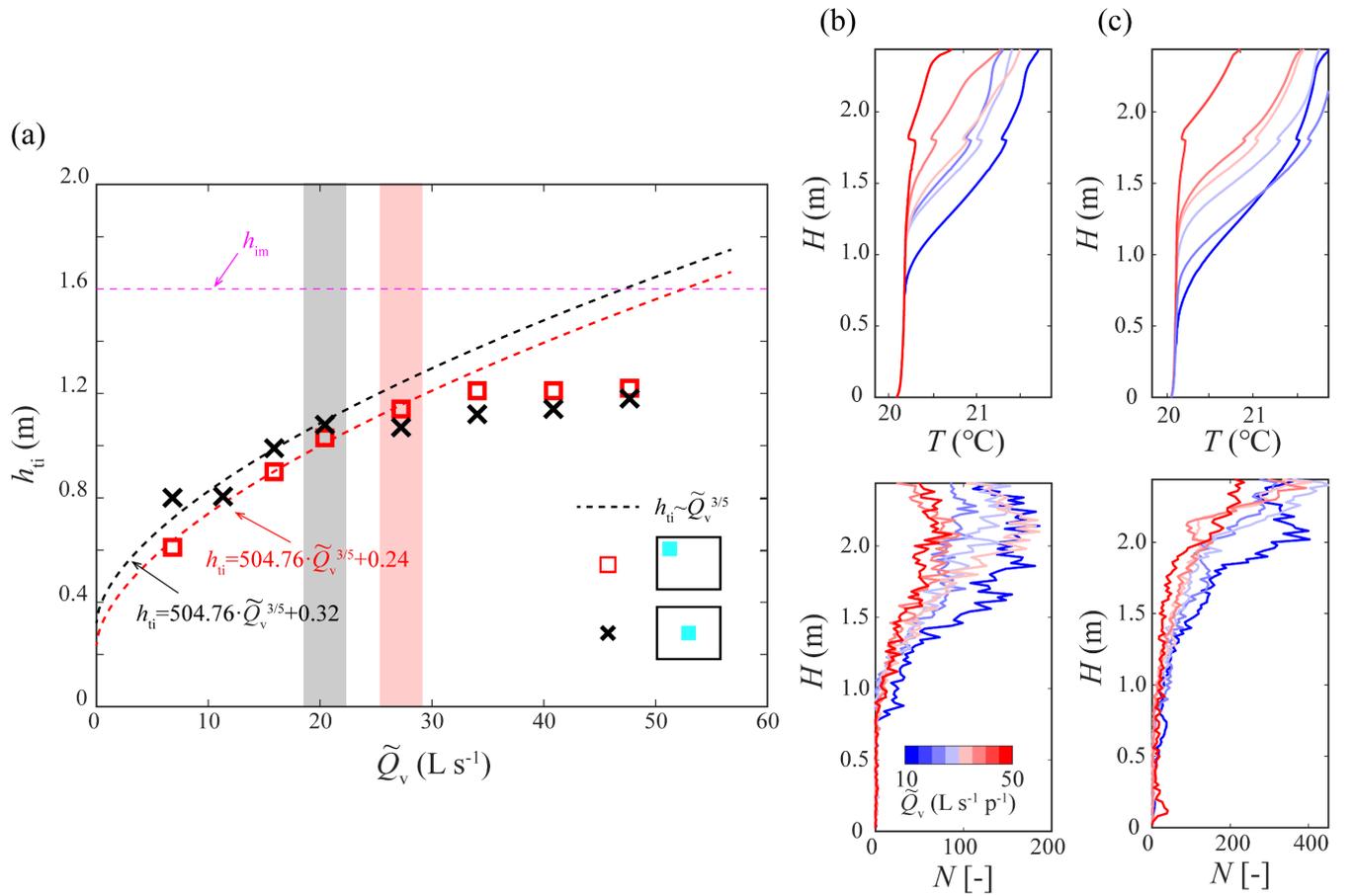

FIG. 11 The variation of (*a*) thermal interface height ($h_{ti}$) with ventilation rate per person ($\widetilde{Q_v}$) for both center- and corner-located infector cases along with the theoretical prediction by classical $h_{ti} \sim \widetilde{Q_v}^{3/5}$ [36, 37, 38], and (*b*, *c*) temperature profile (top row) and particle number stratification along the vertical direction (bottom row) under different ventilation rates for (b) center- and (c) corner-located infector cases, respectively.

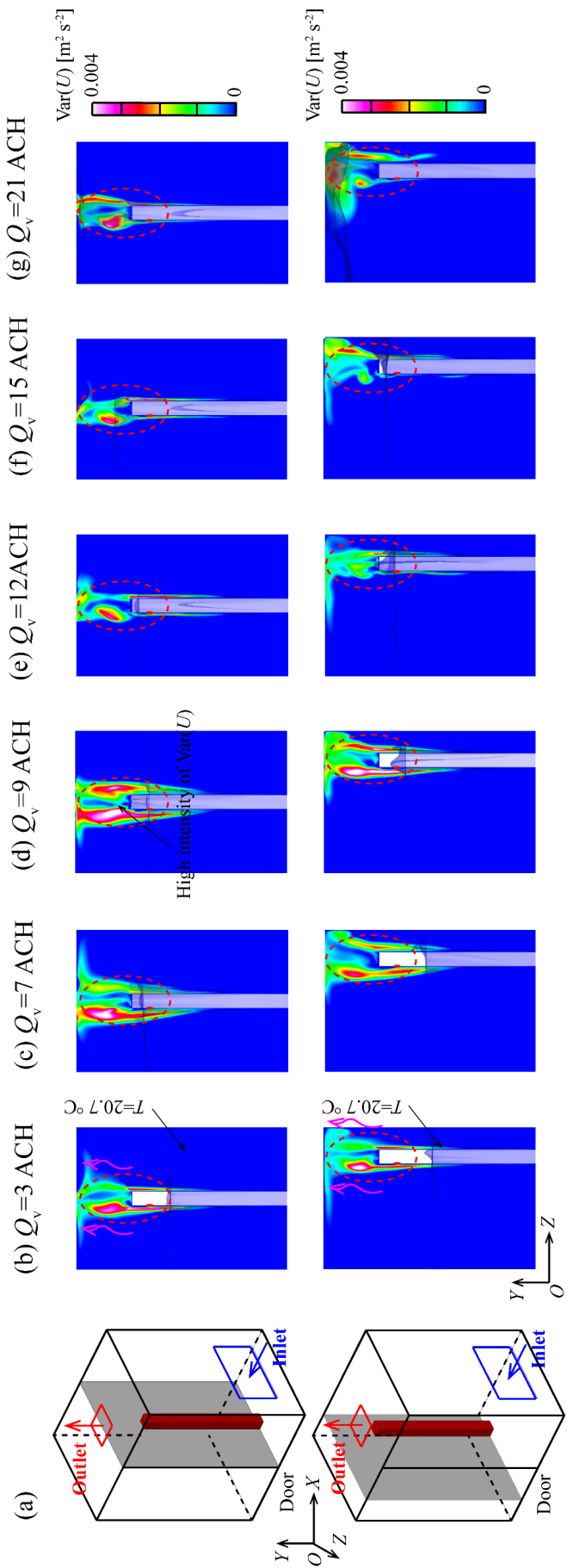

FIG. 12 Influence of ventilation rate ($Q_v$) on variance of velocity fluctuations, i.e., Var($U$), along with the temperature isosurface of $T$=20.7 °C on the middle plane crossing the middle of infector in the model room for center- (top row) and corner-located (bottom row) infectors as illustrated in the schematic of cases of the model room where arrows indicate the ventilation inlet and outlet in ($a$), for ($b$) $Q_v$=3 ACH, ($c$) $Q_v$=7 ACH, ($d$) $Q_v$=9 ACH, ($e$) $Q_v$=12 ACH, ($f$) $Q_v$=15 ACH, and ($g$) $Q_v$=21 ACH, respectively.

Particle dispersion and its removal efficiency in indoor environments are closely related to the infection risk of airborne transmission. To quantitatively analyze the thermally stratified flow (using temperature field) and aerosol particle dispersion in the breathing zone concerning the infection risk of airborne transmission, the locally mean temperature along with its standard deviation and particle number percentage ($\beta$) from ventilation rate $Q_v$=3 ACH to 21 ACH for both center- and corner-located infector cases are presented in Fig. 13 and Fig. 14, respectively. Here, $\beta$ is calculated based on the percentage of the number of particles suspended in the local region in the breathing zone to the number of total particles injected in the model room. Generally, both the mean temperature and $\beta$ decrease with increasing $Q_v$ and approach their constant value at a high ventilation rate. Specifically, for the center-located infector, the mean temperature at the center region of the model room where the infector stands has a higher value, while the mean temperature at the four corners stays at nearly the same low values. In Fig. 13, we find that the mean temperature in the whole breathing zone is lower than that at the center of the model room and just slightly higher than that at the corners. With increasing $Q_v$, the temperature first increases ($Q_v$=3 ACH ~ 5 ACH) and then decreases ($Q_v$=5 ACH ~ 21 ACH). For $\beta$ in Fig. 14, two regions of high $\beta$ values are observed, including the right and left front corners, with lower $\beta$ values at the center, left and right back corners of the room. The aerosol transmission from the room center to the front corners by the infector shows the possibility of local hot spots of infection risks even far away from the infector due to the aerosol transmission caused by ventilation and thermal plumes. These local hot spots associated with infector location can also be identified from both local temperature distribution and $\beta$ for the corner-located infector in the breathing zone, while two regions of high $\beta$ value are located at the right and left back corners where the infector stands.

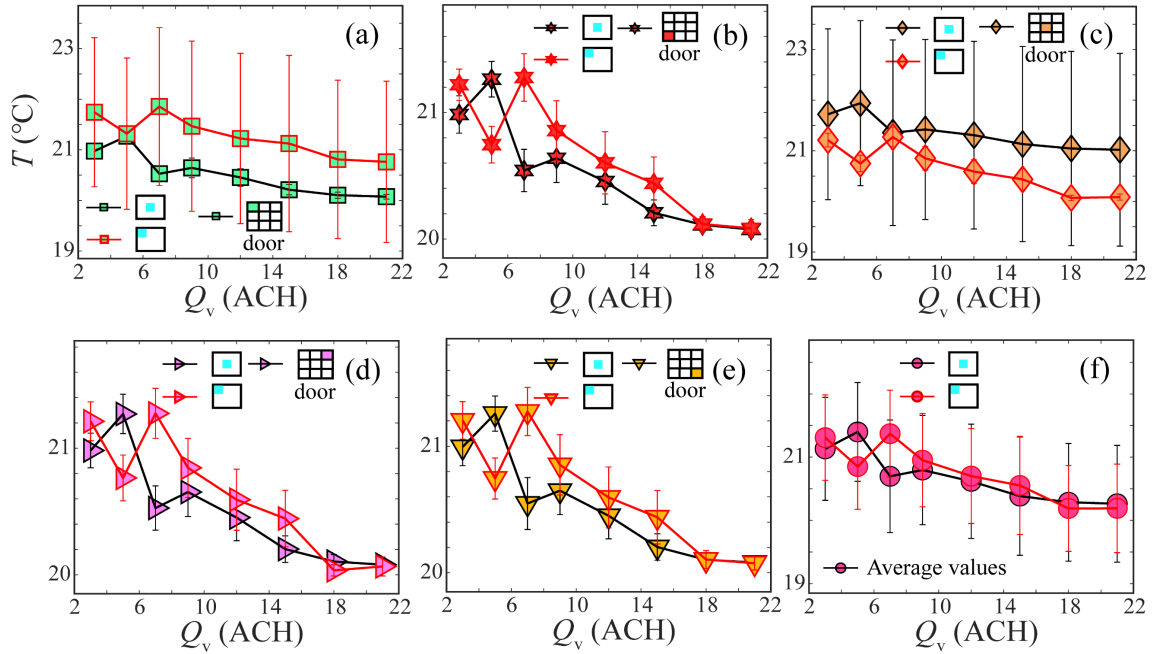

FIG. 13 Local distribution of mean temperature along with its standard deviation shown in error-bars in the breathing zone ($H$=1.40 m~1.80 m) for both center- (black lines) and corner-located infector cases (red lines) for (*a*) at the back left corner, (*b*) at the front left corner, (*c*) at the center, (*d*) at the back right corner, (*e*) at the front right corner, and (*f*) average temperature in the breathing zone.

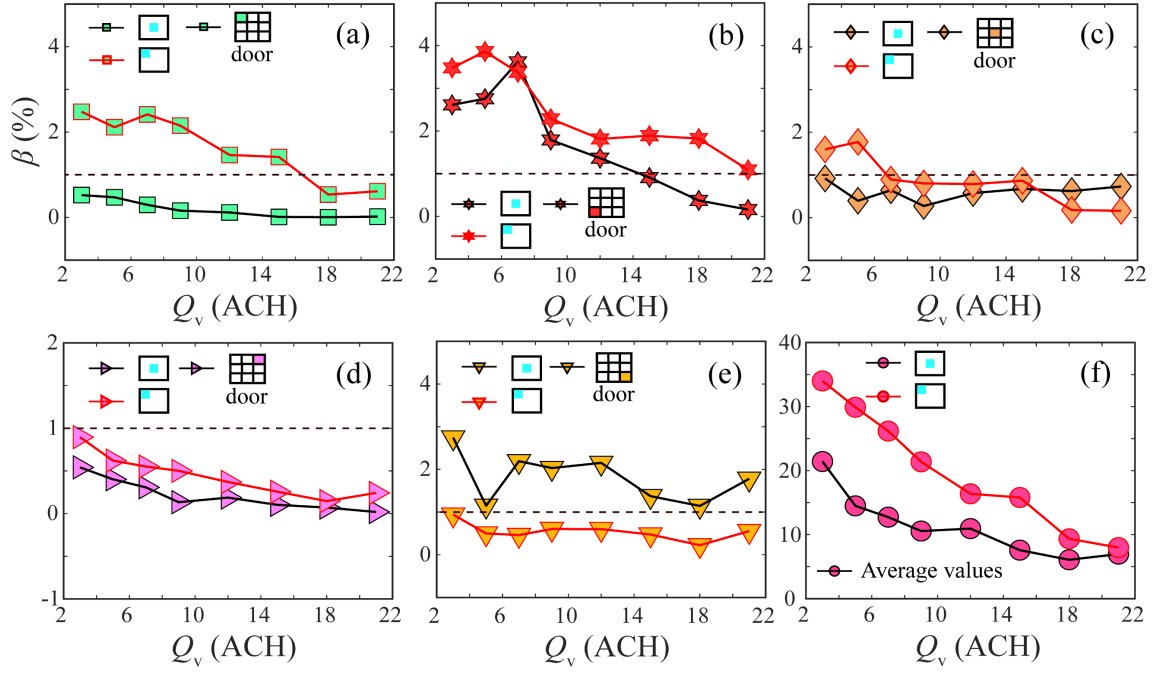

FIG. 14 Locally distribution of particle number percentage in terms of total particle number injected in the breathing zone ($H$=1.40 m~1.80 m) for both center- (black lines) and corner-located infector cases (red lines) for (*a*) at the back left corner, (*b*) at the front left corner, (*c*) at the center, (*d*) at the back right corner, (*e*) at the front right corner, and (*f*) total particle number percentage in the breathing zone.

## B. Effect of ventilation temperature on aerosol particle dispersion

From the above, we find that the infection risk of airborne transmission in a low-ceiling room under displacement ventilation is closely associated with the thermal stratification fields, i.e., $h_{ti}$, showing the importance of the thermal environments on particle dispersion and its removal efficiency, thus the risk of airborne infection. Below $Q_c$, increasing $Q_v$ will significantly increase $\varepsilon_p$, thus reducing the infection risk of airborne transmission. However, above $Q_c$, further increasing $Q_v$ only causes a slower increase rate of $\varepsilon_p$, where more energy is required to achieve a higher $\varepsilon_p$, which would not be efficient in terms of reducing infection risk. Instead, several other mitigation measures, such as air filtration and optimal placement of ventilation settings (i.e., inlet and outlet), can be employed to aid the reduction of infection risk. Here, we conduct studies to investigate the influence of ventilation temperature on $\varepsilon_p$ and propose a thermal-based mitigation strategy by changing the ventilation temperature to alter/disturb the thermal stratification fields, thus controlling the particle dispersion and its removal efficiency, i.e., eventually helping mitigate the infection risk in indoor environments. To study the influence of ventilation thermal effects on airborne transmission (i.e., $\varepsilon_p$) and provide guidance for thermal-based airborne transmission mitigation strategies, we conducted additional simulations by increasing and decreasing the ventilation temperatures while keeping the ambient room temperature constant. The ventilation conditions chosen are near $Q_c$, i.e., slightly lower than $Q_c$ for the center-located infector and slightly higher than $Q_c$ for the corner-located infector, under different infector locations for both the center- (Case T1 and Case T3) and corner-located (Case T2 and Case T4) infector cases.

Fig. 15 presents the comparisons of $\varepsilon_p$ at different ventilation temperature conditions. For the center-located infector case, $\varepsilon_p$ at the high ventilation temperature ($T$=21 °C) of approximately 64.7% is much higher than that at the low ventilation temperature ($T$=19 °C) approximately 40.7%, while both of them are lower than that at $T$=20 °C of approximately 68.4%. For the corner-located infector, $\varepsilon_p$ at the high ventilation temperature ($T$=21 °C) of approximately 47.3% is only slightly higher than that at the low ventilation temperature ($T$=19 °C) of approximately 46.0%, and both are quite similar to that at $T$=20 °C of approximately 47.7%. Consequently, $\varepsilon_p$ for the center-located infector where the ventilation flow directly interacts with body plumes is more sensitive to the ventilation temperature effect than that for the corner-located infector. Moreover, increasing the ventilation temperature ($\Delta T$=1 °C) shows a higher $\varepsilon_p$ than decreasing the ventilation temperature ($\Delta T$=-1 °C). In the following, we inspect the thermal stratification and flow fields under different ventilation temperature conditions for center- and corner-located infectors to probe into the involved physics.

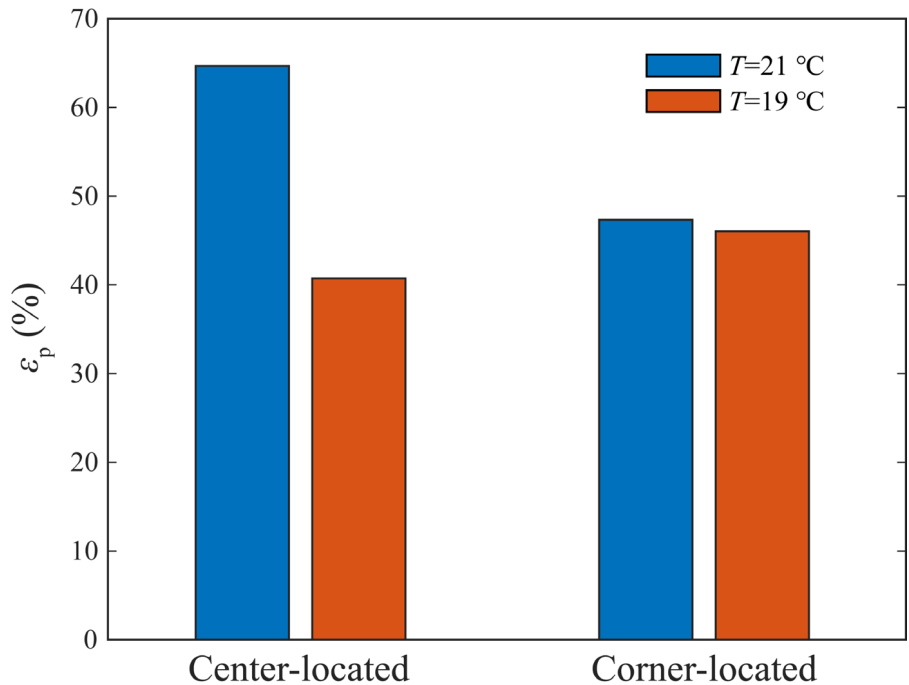

FIG. 15 Comparisons of particle removal efficiency under two ventilation temperature settings, i.e., $T$=19 °C and $T$=21 °C, for both center- ($Q_v$=7 ACH) and corner-located infectors ($Q_v$=15 ACH). Note that the ambient temperature ($T$=20 °C) in the model room stays.

To quantitatively analyze the physics that lead to the increase (or decrease) of $\varepsilon_p$ when increasing or decreasing ventilation temperature, Fig. 16 and Fig. 17 show the stratification fields of temperature and velocity on the $x$-$y$ plane and $y$-$z$ plane crossing the center of the infector in the model room under the conditions corresponding to those in Fig. 15. We find that the temperature stratification fields can be significantly modified when the ventilation temperature is higher or lower than the ambient temperature, i.e., ± 1 °C. For the center-located infector case in Fig. 16, where the infector body could have strong blocking effects on the ventilation flows, at a lower ventilation temperature $T$=19 °C, the cooler ventilated air generally goes down, resulting in the expansion of both the lower cool layer and the upper warm layer toward the floor. As a consequence, the temperature gradient within the region below the infector height decreases. However, $h_{T,20.7}$, the relative height with respect to the infector mouth height that is significant to $\varepsilon_{p,c}$, as illustrated in Fig. 5 and Fig. 7, decreases slightly. Accordingly, we also observe that $\varepsilon_p$ lowers slightly in Fig. 15. Moreover, the separated region above the infector head with low temperature and velocity expands due to the entrainment of a large amount of ventilated cooler air, and the strength of the vertical velocity stream decreases, as marked by the red dashed ellipses in Fig. 16 (a). At the same time, the interface of the lower cool layer becomes wavy, as shown in Fig. 16 (a), due to the interactions between the cooler ventilation flow and the cool clear region. At a higher ventilation temperature $T$=21 °C in Fig. 16 (b), the warmer ventilated air goes upward and separates the original thermal stratification fields, resulting in the warm layer (> 21 °C) going upward and the cool layer (< 21 °C) going downward. Consequently, the layer with a similar temperature (~21 °C) becomes wider, and the cool region shrinks, as shown in the temperature fields in both the $x$-$y$ plane and $y$-$z$ plane in Fig. 16 (b). Moreover, $h_{ti}$ is much lower than that at $T$=19 °C and $T$=20 °C, leading to a smaller particle removal efficiency, as observed in Fig. 16. In Fig. 16 (b), the temperature layer for $T$=21 °C with the same temperature as the ventilation temperature becomes wider. The temperature in the region above the infector in the high ventilation temperature case ($T$=21 °C) is higher than that in the low ventilation temperature case ($T$=19 °C), and a larger region of low temperature and low velocity above the infector head is observed, as indicated by the dashed red ellipses, where more particles can be trapped, thus reducing the particle removal efficiency.

For the corner-located infector case in Fig. 17, where the influence of the infector body on the ventilation flow is weaker, the thermal stratification flow fields change mainly due to the interactions between the ventilation flow and thermal stratification flow fields for low and high ventilation temperature conditions illustrated in Fig. 17 (a) and Fig. 17 (b), respectively. At low temperature ($T$=19 °C), the low ventilated air goes downward, lowering the cool air layer and causing the interface between the cool and warm air layers to oscillate. Accordingly, the temperature gradient across the interface decreases; however, $h_{ti}$, which is closely related to the particle removal efficiency remains nearly the same. At higher temperatures ($T$=21 °C), the interface lowers further, while the temperature layer at $T$=21 °C becomes wider. The separated region above the infector head enlarges significantly, where more particles can be trapped there, reducing the particle removal efficiency, which is the same as that for the center-located infector. It is worth noting that although the flow fields are altered at high ventilation temperatures, the influence of ventilation temperature on the particle removal efficiency is relatively small, where particle removal efficiency increases slightly at high ventilation temperatures. Therefore, when changing the ventilation temperature, the thermal stratification flow fields change considerably due to the interactions between the ventilation flow and the specified temperature layer and the block effects of the infector body, thus potentially changing the corresponding $\varepsilon_p$. In Fig. 18, the profiles of mean temperature and particle percentage are presented to show the thermal stratification fields and particle dispersion along the vertical direction. For both the center- and corner-located infectors, the temperature profiles for high ventilation temperatures show a lower temperature gradient, especially in the breathing zone, and the particle percentage profiles show a smaller value near the room ceiling. Consequently, increasing the ventilation temperature ($\Delta T$=1 °C) is better than decreasing the ventilation temperature ($\Delta T$=-1 °C) in terms of $\varepsilon_p$.

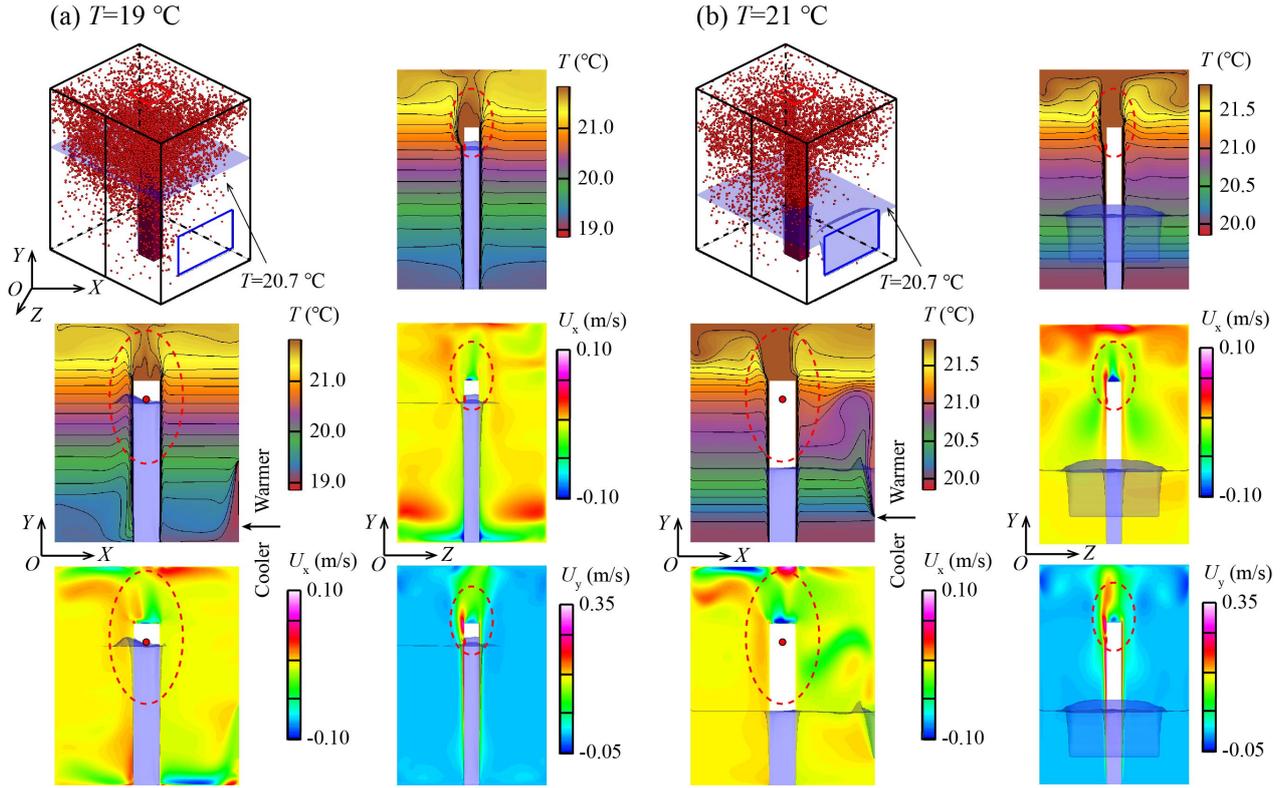

FIG. 16 Thermal stratification (i.e., mean temperature) and flow fields (i.e., mean streamwise and vertical velocity) on the x-y plane corresponding to those in Fig. 5 and the y-z plane corresponding to those in Fig. 7 under two ventilation temperatures for the center-located infector at (a) T=19 ℃ and (b) T=21 ℃.

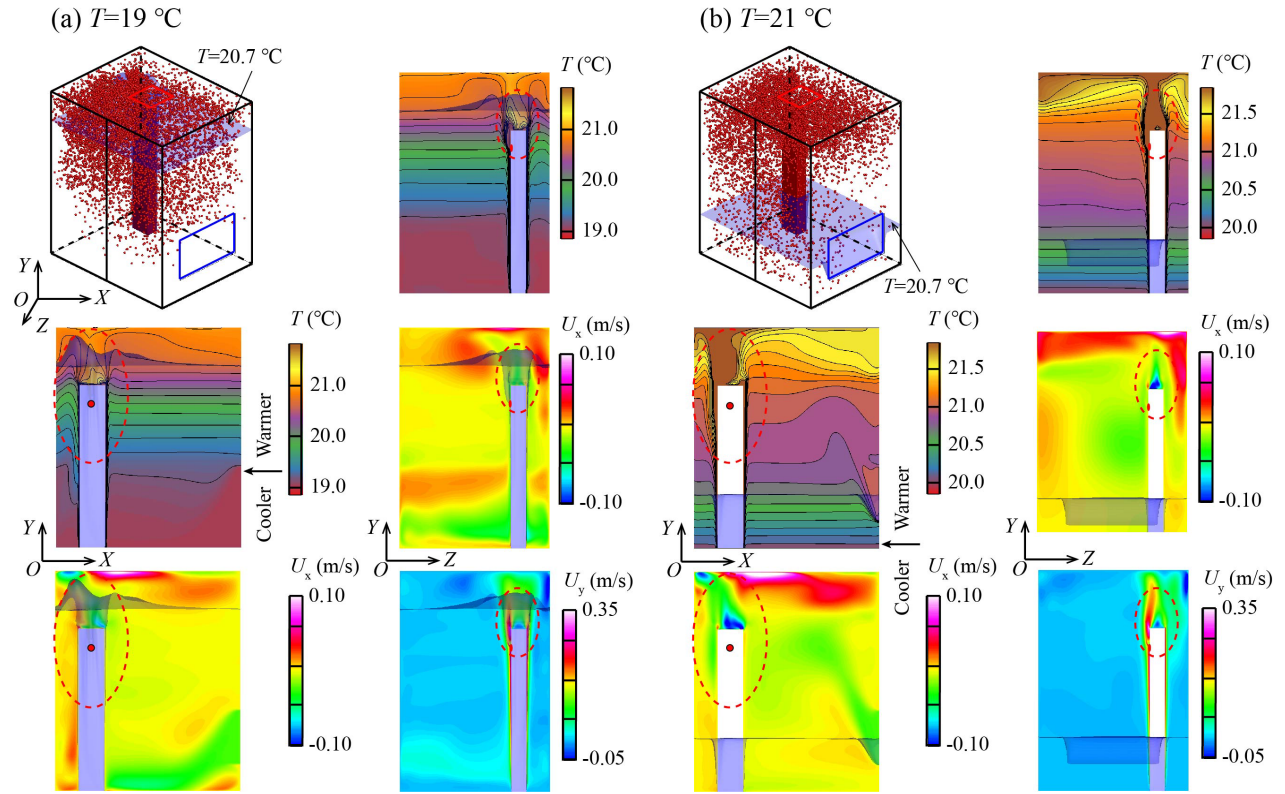

FIG. 17 Thermal stratification (i.e., mean temperature) and flow fields (i.e., mean streamwise and vertical velocity) on the x-y plane corresponding to those in Fig. 8 and the y-z plane corresponding to those in Fig. 10 under two ventilation temperatures for the corner-located infector at (a) T=19 ℃ and (b) T=21 ℃.

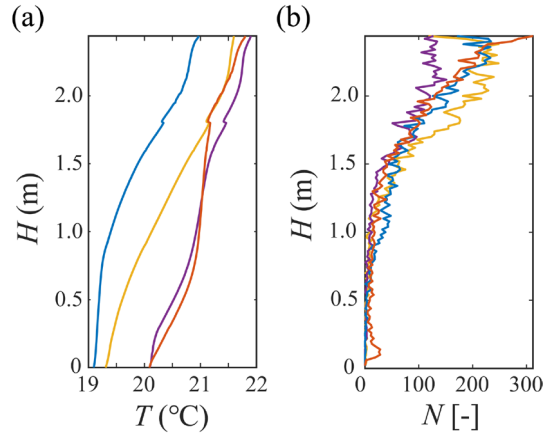

Fig. 18 Temperature profile (a, left) and particle number stratification along the vertical direction (b, right) under varying ventilation temperature settings corresponding to those in Fig. 15. ⎯ Center-located infector at 19 ℃, ⎯ Center-located infector at 21 ℃, ⎯ Corner-located infector at 19 ℃, and ⎯ Corner-located infector at 21 ℃.

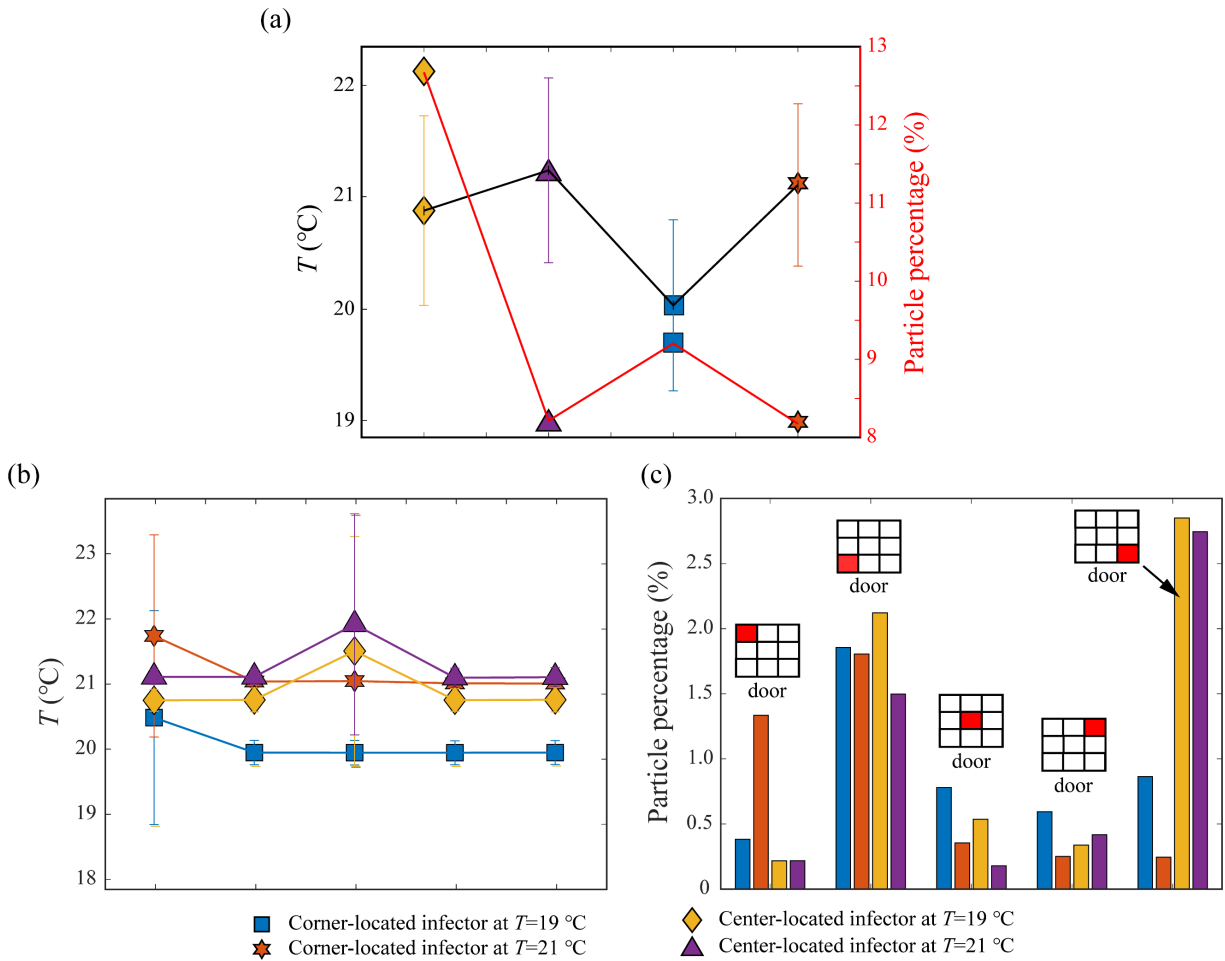

Fig. 19 Distribution of mean temperature and particle percentage averaged in the breathing zone (a) and locally mean temperature (b) and locally particle percentage (c) inside the breathing zone for ventilation temperature settings corresponding to those in Fig. 15.

The mean temperature and particle percentage averaged over the entire breathing zone ($H$=1.40 m~1.80 m) and locally (i.e., at the center and all four corners in the breathing zone) are shown in Fig. 19 in order to quantitatively analyze the flow fields in the breathing zone associated with the risk of infector. For the infection risk in the breathing zone, shown in Fig. 19 (a), the high ventilation temperate conditions show a higher mean temperature for both center- and corner-located infectors and a lower particle percentage, indicating a lower risk of airborne infection. The total particle percentage is less than 15.0% in the breathing zone, with a maximum value at the corner-located infector and a low temperature case of 12.7%. For the risk assessment for locally mean flow fields in the breathing zone, shown in Fig. 19 (b) and Fig. 19 (c), the locally mean temperature near the infector shows high values, while the other local regions show almost similar values. The influence of the infector location and ventilation inlet/outlets on the local particle percentage is significant. Specifically, the locations in front of the infector show a higher particle percentage, where all four ventilation temperature cases for both center- and corner-located infectors show a higher particle percentage in front of the infector both on the same side and opposite side of the ventilation inlet, while locations behind the infector show relatively lower particle percentages, resulting in a lower risk of airborne infection. It is worth noting that although the region near ventilation inlet shows low particle percentage for corner-located infector, the highest particle percentage occurs at the right corner near the ventilation inlet in front of the infector for the center-located infector, which is due to the block effects of infector body to the ventilation flows. Consequently, it is suggested that when standing inside an enclosed indoor environment such as an elevator cabin, people should avoid standing near the ventilation inlet to block the ventilation, which could potentially result in local hot pots.

## IV. CONCLUSIONS AND DISCUSSION

Using computational fluid dynamics (CFD), this study presents a systematic numerical investigation of the risk of airborne disease transmission, such as SARS-CoV-2, in a low-ceiling room with a displacement ventilation system and a possible temperature-based mitigation strategy. A test model room with low-ceiling, which has the same size as a 3,500-pound elevator cabin, is established with one occupant standing inside, representing a passenger taking the elevator. The displacement ventilation system has one inlet located at the lower part of the room sidewall and an outlet located at the room ceiling center. The numerical simulations are conducted using the OpenFOAM C++ libraries. We employed the Euler-Lagrange computation approach to calculate the air flows generated by both the ventilation and buoyancy heat source (i.e., infector) and the corresponding aerosol particle dispersion. Our study shows that due to the confinement effects in a low-ceiling room, the thermal stratification lock-up effects, which are a unique nature of displacement ventilation, are significant where the infector could stay both in the lower cool region and higher warm region crossing the thermal stratification interface, and the thermal interface height is lower than that in a tall-ceiling room. With increasing ventilation rates, first, the lock-up interface height increases and then approaches its maximum and then further increases the ventilation rate, and the thermal interface height increases in a much slower rate or even remains constant. Subsequently, strong interactions exist between ventilation flows, buoyancy flows, and respiratory flows. Furthermore, we find that the infector location with respect to ventilation inlet and outlet has an important influence on the thermal interface height. Compared with that in the center-located infector case, the thermal interface height is higher at the corner-located infector. However, for both the center- and corner-located infector, at a low ventilation rate range, the relationship between the thermal interface height and ventilation rate satisfies the classical 3/5 law [36], while the deviation begins at the transition ventilation rate.

By measuring the particle removal efficiency and spatial particle number distribution, we find that variations in particle removal efficiency and particle number distribution with ventilation rates show a similar trend as the thermal stratification interface height, showing a close relationship between the particle removal efficiency and thermal stratification fields (i.e., thermal interface height). Specifically, in the low ventilation rate range, with increasing ventilation rate, the particle removal efficiency increases, while in the high ventilation rate range above the transition ventilation rate, the particle removal efficiency begins to decrease. Therefore, in terms of the risk of airborne infection, there exists an optimal ventilation rate where the risk of airborne infection is the lowest, above which further increasing the ventilation rate does not reduce the infectious risk efficiently. This phenomenon could be explained by the 'short-circuiting' mechanism [58], where some of the ventilation flow goes directly to the outlet without being entrained into the upper layer via thermal buoyancy flows, and further increasing the thermal interface height and particle removal efficiency could be difficult. Moreover, this transition ventilation rate could be influenced by several factors, i.e., vent locations and size, infector number and locations. In the current study, above the optimal ventilation rate, with increasing ventilation rate, differences in flow fields are also observed between different infector locations. Specifically, for the center-located infector, the particle removal efficiency decreases quickly, while for the corner-located infector, the particle removal efficiency decreases gradually or remains almost constant. By examining the flow fields, we find that the turbulent flow fields above the body in the warm layer are responsible for this phenomenon. For the center-located infector, the standard deviation of velocity, Var($U$), increases with increasing ventilation rate, while for the corner-located infector, the Var($U$) decreases. Our simulation results show a close relationship between the thermal stratification fields and particle removal efficiency, which provides the possibility to mitigate the risk of airborne infection by changing the thermal stratification fields. Finally, we further study the efficiency of thermal mitigation strategies by increasing and decreasing the ventilation temperature ($\Delta T=\pm 1$ °C) while keeping the ambient temperature constant. We find that compared with that with a lower ventilation temperature ($\Delta T=-1$ °C), the ventilation design with a higher ventilation temperature ($\Delta T=+1$ °C) has a higher particle removal efficiency for center-located infector. The thermal stratification fields may be severely affected by ventilation flows, and when the ventilation temperature differs from ambient temperature, the stratified layer with the same temperature as that of ventilation widens. In particular, for the lower ventilation temperature ($\Delta T=-1$ °C), the cooler layer height becomes higher and the interface is wavy, which is caused by the interactions between the ventilation flows and cooler layer, while for the higher ventilation temperature ($\Delta T=+1$ °C), the cooler layer height is reduced with the layer with the same temperature as the ventilation temperature broadens. Clearly, properly adjusting the ventilation temperature can improve the particle removal efficiency, suggesting a potential mitigation strategy of airborne transmission based on internal thermal control. Nevertheless, optimal thermal control for risk mitigation may depend on specific room configuration and occupant locations.

According to our work, the risk of airborne transmission in low-ceiling rooms with strong confinement effects could be more serious than that in high-ceiling rooms, where the strong interactions among ventilation flows, buoyancy flows, and respiratory flows generated by the human body are significant, especially the thermally stratified lock-up effects. Compared to a high-ceiling room (a ceiling height of 3 m with a 1.2 m seated occupant and $Q_c=50$ L/s [54]), the current low-ceiling room (a ceiling height of 2.44m with a 1.8m standing infector) shows a significant decrease in $Q_c$, specifically 20.40 L/s and 27.20 L/s for infectors located in the center and corner, respectively. A lower $Q_c$ in low-ceiling rooms suggests that particle removal efficiency or contaminant concentration reduction begins to level off at a smaller range of ventilation rates, and further increasing ventilation rate has limited impact on reducing virus-laden aerosols or contaminant concentration. For instance, the first case of COVID-19 transmission inside an elevator in South Korea, where two people were together for only a minute without wearing masks, highlights the high risk of infection in small confined spaces with low-ceilings. The study by Escombe et al. [19] suggested that at the same ACH value, rooms with large windows and high ceilings would have greater absolute ventilation than those with small windows and low ceilings, resulting in a lower risk of airborne infection. It is worth noting that in our flow setup, simplified body shapes and inlet and outlet ventilation vents are used. The transition ventilation rate could be influenced by vent locations and size, body posture, infector locations, room size and geometry. Particle removal efficiency may depend on the local flow fields around the realistic body (i.e., buoyancy flows, respiratory flows); however, for a 3-minute simulation, the particle removal efficiency is primarily influenced by thermal stratification fields, while the effects of the current simplified body shape are relatively small. Our work shows that particle dispersion, i.e., particle removal efficiency, is primarily influenced by thermal stratification fields, and consequently, strategies that can alter the thermal stratification fields will certainly modify aerosol particle dispersion and thus the risk of infection. We propose a thermal-based mitigation strategy by adjusting the ventilation temperature. Our simulation shows that both increasing and decreasing the temperature by $\Delta T = \pm 1$ °C result in poorer particle removal efficiency compared to keeping the same temperature as the ambient temperature. Consequently, optimal ventilation thermal control, such as adaptive ventilation based on local parameters like temperature and velocity, is necessary. Future work could also be conducted to systematically investigate the influence of ceiling height (low-ceiling, middle-ceiling, and high-ceiling) on thermal stratification fields and thus the particle removal efficiency. The location of the infector locations (i.e., heat and aerosol source) has a significant impact on the effects of ventilation on particle removal efficiency. It is noteworthy that other factors, such as humidity and evaporation, which are also important for aerosol transmission in indoor environments, as reported by [77], also require further investigation.

## Appendix

To verify the convergence of the current computational mesh, Fig. A1 presents comparisons of the temperature distribution and respiratory aerosol particles dispersion on the y-z plane crossing the center of the infector body at $Q_c$ for center-located infector obtained from the current mesh and further refined mesh. The temperature fields, share the same stratification flow structures, with the $h_{T,20.7}$ at almost the same height, and particles dispersions show the similar locations, as those obtained from the further refined mesh, indicating the convergence of the mesh in Fig. A1 (a, b). A comparison of the mean temperature and particle numbers Fig. A1 (c, d) along the height direction further quantitatively verifies the convergence of our results.

Fig. A2 shows the probability density function of particle diameters in the current simulation. The particle diameter is mainly approximately 2 $\mu$m, below 10 $\mu$m. Therefore, aerosol particles in the current study mainly remain airborne, and few of them deposit on walls for a relatively long-term simulation (i.e., 3 minutes).

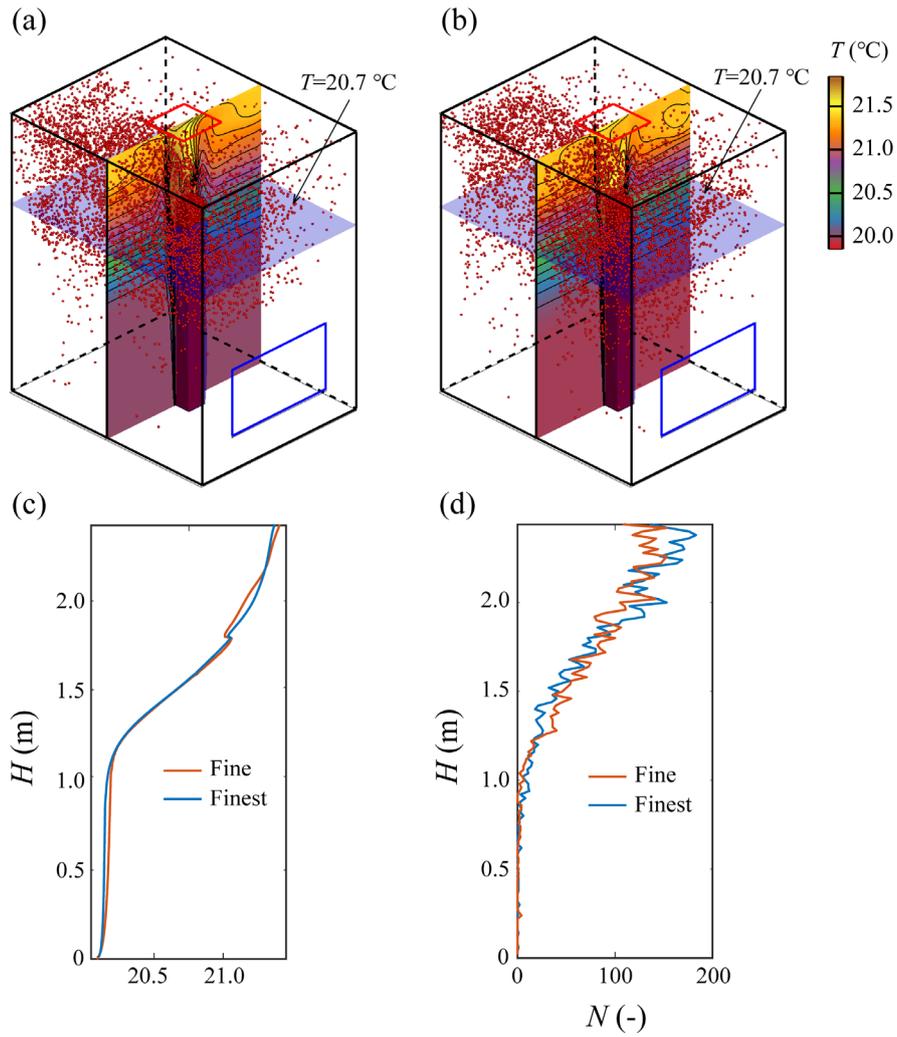

FIG. A1 Comparisons of the temperature distribution and particle dispersions for (a) the grid used in the present study, 3.2×10⁶, and (b) the further refined grid, 3.6×10⁶, and the profiles of (c) temperature and (d) particle number along the room height direction.

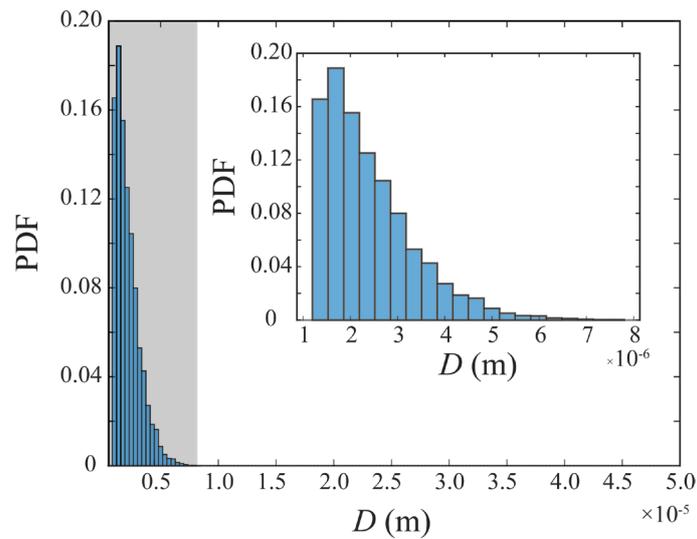

FIG. A2 Probability density function (PDF) of particle diameters used in the current study. The shadow region is enlarged in the insert figure.

As shown in the methodology section, to ensure the statistical convergence of particle trajectories, the particle number used is 10 times from normal breathing activities, and thus 440 particles per breathing cycle. To further validate the convergence of particle trajectories, we conduct simulations using more particles, i.e., $N_T$=100, 200 and 300 times from normal breathing activities. Fig. A3 presents the variations of $\varepsilon_p$ as a function of $N$, by increasing particle number emission rate to 10, 100, 200 and 300, respectively. We find that $\varepsilon_p$ keeps almost the constant even when the particle number emission rate is 300 times from normal breathing activities, indicating the convergence of particle trajectories.

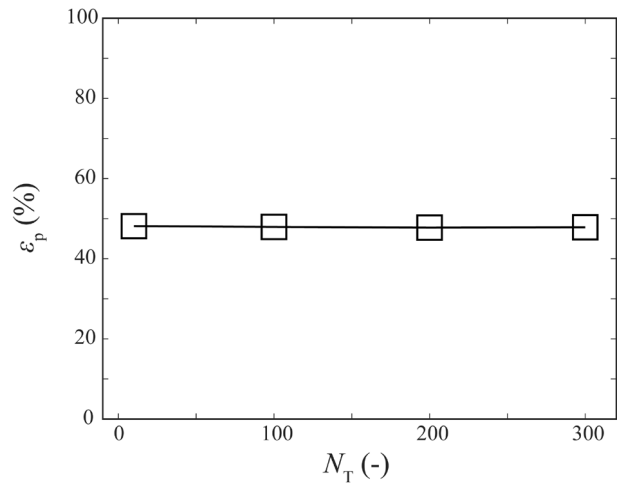

FIG. A3 Convergence study of the effects of particle number emission rate ($N_T$=10, 100, 200, 300, where $N$=current particle number emission rate/real particle number emission rate) on particle removal efficiency ($\varepsilon_p$). $Q_v$=12 ACH and $T$=19°C at corner-located infector.

## Acknowledgment

Financial support of the China Scholarship Council (CSC, Grant No.: 201906030144) is appreciated.